\documentclass{article}
\usepackage[left=2cm, right=2cm, top=2cm, bottom=2cm]{geometry}
\usepackage{graphicx}
\usepackage{authblk}
\usepackage{amsmath}
\usepackage{mathrsfs}
\usepackage{amsfonts}
\usepackage{gensymb}
\usepackage{hyperref}
\usepackage{MathUnicode}

\newcommand{\F}{\mathcal{F}}
\newcommand{\PSF}{\textrm{PSF}}
\renewcommand{\d}{\textrm{d}}
\newcommand{\A}{\mathcal{A}}

\providecommand{\keywords}[1]{\textbf{\textit{Keywords ---}} #1}

\title{Asymmetries in adaptive optics point spread functions}
\author[1]{Alexander Madurowicz}
\author[1]{Bruce Macintosh}
\author[2]{Vanessa P. Bailey}
\author[3]{Jeffrey Chilcote}
\author[4]{Marshall Perrin}
\author[5]{Lisa Poyneer}
\author[4]{Laurent Pueyo}
\author[1]{Jean-Baptiste Ruffio}
\author[6]{Travis Barman}
\author[7]{Joanna Bulger}
\author[8]{Tara Cotten}
\author[1]{Robert J. De Rosa}
\author[9]{Rene Doyon}
\author[10]{Gaspard Duch\^ene}
\author[10]{Thomas M. Esposito}
\author[11]{Michael P. Fitzgerald}
\author[12]{Katherine B. Follette}
\author[13]{Benjamin L. Gerard}
\author[14]{Stephen J. Goodsell}
\author[10]{James R. Graham}
\author[15]{Alexandra Z. Greenbaum}
\author[16]{Pascale Hibon}
\author[17]{Li-Wei Hung}
\author[18]{Patrick Ingraham}
\author[10]{Paul Kalas}
\author[19]{Quinn Konopacky}
\author[19]{J\'er\^ome Maire}
\author[20]{Franck Marchis}
\author[21]{Mark S. Marley}
\author[22]{Christian Marois}
\author[23]{Stanimir Metchev}
\author[2]{Maxwell A. Millar-Blanchaer}
\author[1]{Eric L. Nielsen}
\author[24]{Rebecca Oppenheimer}
\author[5]{David Palmer}
\author[25]{Jennifer Patience}
\author[4]{Abhijith Rajan}
\author[9]{Julien Rameau}
\author[26]{Fredrik T. Rantakyr\"o}
\author[27]{Dmitry Savransky}
\author[4]{Anand Sivaramakrishnan}
\author[8]{Inseok Song}
\author[4]{Remi Soummer}
\author[1]{Melissa Tallis}
\author[18]{Sandrine Thomas}
\author[10]{Jason J. Wang}
\author[25]{Kimberly Ward-Duong}
\author[28]{Schuyler Wolff}

\affil[1]{Kavli Institute for Particle Astrophysics and Cosmology, Stanford University, Stanford, CA 94305, USA}
\affil[2]{Jet Propulsion Laboratory, California Institute of Technology, Pasadena, CA 91109, USA}
\affil[3]{Department of Physics, University of Notre Dame, 225 Nieuwland Science Hall, Notre Dame, IN, 46556, USA}
\affil[4]{Space Telescope Science Institute, Baltimore, MD 21218, USA}
\affil[5]{Lawrence Livermore National Laboratory, Livermore, CA 94551, USA}
\affil[6]{Lunar and Planetary Laboratory, University of Arizona, Tucson AZ 85721, USA}
\affil[7]{Subaru Telescope, NAOJ, 650 North A{'o}hoku Place, Hilo, HI 96720, USA}
\affil[8]{Department of Physics and Astronomy, University of Georgia, Athens, GA 30602, USA}
\affil[9]{Institut de Recherche sur les Exoplan{\`e}tes, D{\'e}partement de Physique, Universit{\'e} de Montr{\'e}al, Montr{\'e}al QC, H3C 3J7, Canada}
\affil[10]{Department of Astronomy, University of California, Berkeley, CA 94720, USA}
\affil[11]{Department of Physics \& Astronomy, University of California, Los Angeles, CA 90095, USA}
\affil[12]{Physics and Astronomy Department, Amherst College, 21 Merrill Science Drive, Amherst, MA 01002, USA}
\affil[13]{University of Victoria, 3800 Finnerty Rd, Victoria, BC, V8P 5C2, Canada}
\affil[14]{Gemini Observatory, 670 N. A'ohoku Place, Hilo, HI 96720, USA}
\affil[15]{Department of Astronomy, University of Michigan, Ann Arbor, MI 48109, USA}
\affil[16]{European Southern Observatory, Alonso de Cordova 3107, Vitacura, Santiago, Chile}
\affil[17]{Natural Sounds and Night Skies Division, National Park Service, Fort Collins, CO 80525, USA}
\affil[18]{Large Synoptic Survey Telescope, 950N Cherry Ave., Tucson, AZ 85719, USA}
\affil[19]{Center for Astrophysics and Space Science, University of California San Diego, La Jolla, CA 92093, USA}
\affil[20]{SETI Institute, Carl Sagan Center, 189 Bernardo Ave.,  Mountain View CA 94043, USA}
\affil[21]{NASA Ames Research Center, Mountain View, CA 94035, USA}
\affil[22]{National Research Council of Canada Herzberg, 5071 West Saanich Rd, Victoria, BC, V9E 2E7, Canada}
\affil[23]{Department of Physics and Astronomy, Centre for Planetary Science and Exploration, The University of Western Ontario, London, ON N6A 3K7, Canada}
\affil[24]{Department of Astrophysics, American Museum of Natural History, New York, NY 10024, USA}
\affil[25]{School of Earth and Space Exploration, Arizona State University, PO Box 871404, Tempe, AZ 85287, USA}
\affil[26]{Gemini Observatory, Casilla 603, La Serena, Chile}
\affil[27]{Sibley School of Mechanical and Aerospace Engineering, Cornell University, Ithaca, NY 14853, USA}
\affil[28]{Leiden Observatory, Leiden University, P.O. Box 9513, 2300 RA Leiden, The Netherlands}

\begin{document}
\maketitle

\begin{abstract}
    An explanation for the origin of asymmetry along the preferential axis of the PSF of an AO system is developed. When phase errors from high altitude turbulence scintillate due to Fresnel propagation, wavefront amplitude errors may be spatially offset from residual phase errors. These correlated errors appear as asymmetry in the image plane under the Fraunhofer condition. In an analytic model with an open-loop AO system, the strength of the asymmetry is calculated for a single mode of phase aberration, which generalizes to two dimensions under a Fourier decomposition of the complex illumination. Other parameters included are the spatial offset of the AO correction, which is the wind velocity in the frozen flow regime multiplied by the effective AO time delay, and propagation distance or altitude of the turbulent layer. In this model, the asymmetry is strongest when the wind is slow and nearest to the coronagraphic mask when the turbulent layer is far away, such as when the telescope is pointing low towards the horizon. A great emphasis is made about the fact that the brighter asymmetric lobe of the PSF points in the opposite direction as the wind, which is consistent analytically with the clarification that the image plane electric field distribution is actually the inverse Fourier transform of the aperture plane. Validation of this understanding is made with observations taken from the Gemini Planet Imager, as well as being reproducible in end-to-end AO simulations.
\end{abstract}


\section{Introduction}
The advancement of Adaptive Optics (AO) as a technology has enabled significant progress in astrophysics. Notably, the Gemini Planet Imager is one such instrument \cite{Poyneer16}, where fast and precise correction are pivotal to optimize instrument performance. The results so far have been spectacular. With detections of multiple planetary mass companions around various stars, as well a strong non-detection limits around many more, one can constrain planetary population distributions.\cite{Nielsen19} Astrometric measurements of multi-body systems probe dynamical constraints on planetary masses and system lifetimes \cite{Wang2018}, and provide a spectacular view of Kepler's Laws in action. Spectral measurements of individual giant planets constrain evolutionary and atmospheric models \cite{Rajan2017} of these objects, paving the way towards characterization of extrasolar terrestrial planets.

For this generation of instruments and the next, understanding the Point Spread Function (PSF) of AO instruments on giant telescopes will be important for development of algorithms optimized in the search for planets \cite{Ruffio2017}\cite{Cantalloube2018B}. The analysis in this paper expands on our previous work \cite{Madurowicz2018}, which demonstrated the origin of azimuthal asymmetry in the PSF as a consequence of the time lag error, to explore asymmetry along the preferential axis introduced by scintillation. This effect has been demonstrated previously by Cantalloube et al. 2018 \cite{Cantalloube2018}. We will expand on their discussion by using a more general method of analyzing the structure of the AO-corrected PSF analytically, as well as validating our conclusions with observations and atmospheric datasets. More specifically, our formalism demonstrates that the asymmetry grows linearly only for small spatial frequencies, and at higher spatial frequencies becomes non-linear. We include solutions for the zeros of the log of the asymmetry metric, which are image locations with an observable return to symmetry.

The analysis in the paper is presented as a trident - theory, simulations, and observations. The first section derives the method of angular spectrum Fresnel propagation from the time independent wave equation. This technique allows us to analytically calculate the PSF formed from a single mode of phase aberration which is both scintillated and time-lag corrected. The PSF for this single mode is computed to second order in a Taylor expansion, which is well matched when compared to a numerical solution involving the discrete Fourier transform. In the second section, these methods are extended broadly to waves propagating through an atmospheric model with Kolmogorov turbulence in the frozen flow regime, which reproduces the behavior in a moderately accurate simulation of an entire telescope employing Adaptive Optics. The third section demonstrates the effects from our analytic model is observable in real data taken from the Gemini Planet Imager. We then explore correlations in the observations when combined with a meteorological dataset containing the real wind velocities and directions in the atmosphere during the observations. Finally, the paper concludes with a brief discussion about the importance of this effect in the context of improving Adaptive Optics systems performance from design to post-processing.

\section{Theoretical Scintillation Analysis}
\subsection{Angular Spectrum Fresnel Propagation}
To evaluate the effects of Fresnel propagation and scintillation, we derive the method of the angular spectrum. For an arbitrary complex illumination $U$ of the electric field, we can consider its decomposition into its angular spectrum $\widetilde{U}$ given by the two-dimensional inverse Fourier transform in the plane ($x$, $y$) at some constant $z$
\begin{equation}
    U(x, y, z) = \iint \limits_{-\infty}^{+\infty} \widetilde{U}(k_x, k_y, z) e^{i (k_x x + k_y y)}\textrm{d}k_x\textrm{d}k_y .
\end{equation}
Direct application of the Helmholtz equation
\begin{equation}
    (\nabla^2 + k^2)U(x,y,z) = 0
\end{equation}
to this decomposition will allow us to derive a formula for the propagation through free space of an arbitrary illumination. For a wave propagating with wavevector $\vec{k} = k_x \hat{x} + k_y \hat{y} + k_z\hat{z}$, implying $k^2 = |\vec{k}|^2 = k_x^2 + k_y^2 + k_z^2$, we find that propagation along the z-axis is constrained by the second-order ordinary differential equation
\begin{equation}
    \frac{\textrm{d}^2}{\textrm{d}z^2}\widetilde{U}(z) + k_z^2\widetilde{U}(z) = 0 ,
\end{equation}
where we have implicitly included the dependence of $\widetilde{U}$ on the particular mode $(k_x, k_y)$. This differential equation permits solutions of the form
\begin{equation}
    \widetilde{U}(z) = \widetilde{U}(z = 0) e^{\pm i k_z z},
\end{equation}
where the $\pm$ represents a wave traveling in the positive or negative z direction. For a mode with $k_x$, $k_y$, and the magnitude of the wavevector constrained by $|k| = 2\pi/\lambda$, where $\lambda$ is the wavelength of the propagating light (assumed monochromatic), this means we can find the angular spectrum at some later plane $z$ from the angular spectrum at the origin $z = 0$ by simply multiplying by the free space propagation transfer function $H(z)$, which takes on the form
\begin{equation}
    H(k_x, k_y, z) = \textrm{exp} \Big[ \pm i \sqrt{ \Big(\frac{2\pi}{\lambda}\Big)^2 - (k_x^2 + k_y^2)} z \Big] .
\end{equation}
Making the substitution $k_{x,y} = 2\pi f_{x,y}$ to represent the true linear wavenumber as in Goodman \cite{goodman1996} recovers this expression for the free space propagation transfer function
\begin{equation}
H(f_x, f_y, z) = \textrm{exp} \Big[ \pm i \frac{2\pi}{\lambda}\sqrt{1 - \lambda^2(f_x^2 + f_y^2)} z \Big] .
\end{equation}
This allows us to consider the propagation of light through free space as a linear spatial filter applied to each Fourier mode of the complex illumination independently. 

\subsection{Single Mode Analysis}

\begin{figure}[b!]
    \centering
    \includegraphics[width=\textwidth]{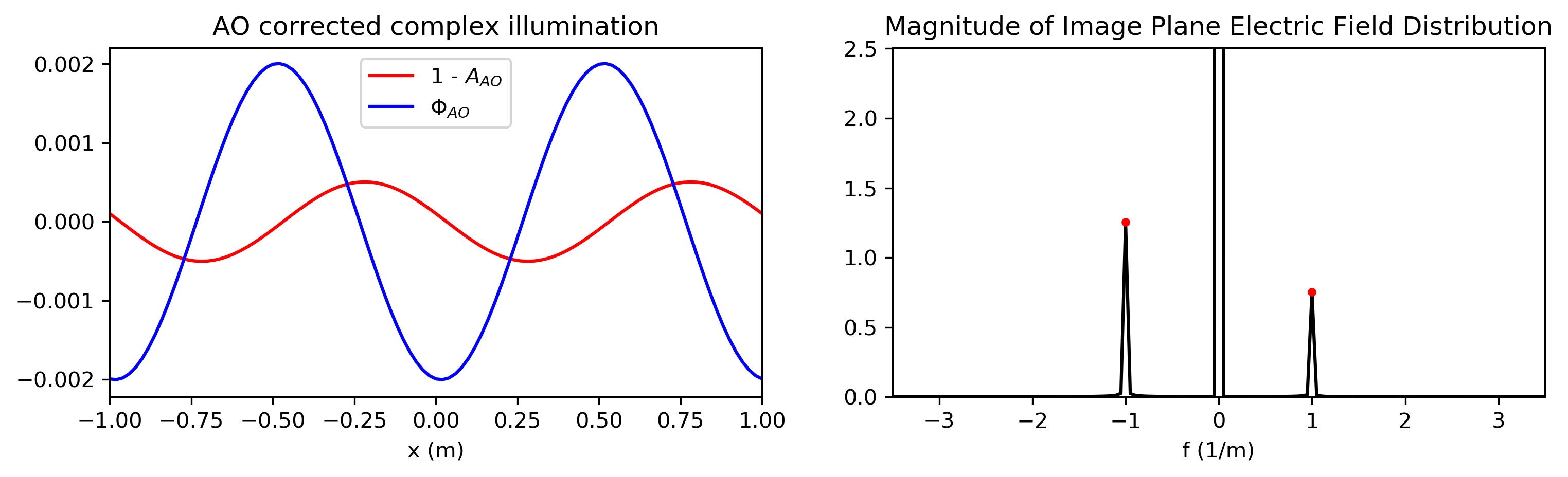}
    \caption{An example of a single mode of AO-corrected complex illumination and corresponding asymmetric speckles in the PSF. Here $\alpha = .01$ rad, $z = 10$ km, $p = 1$ m, $v = 10$ m/s, $t = 3.2$ ms. The AO-corrected complex illumination shows small aberrations in both amplitude and phase from the Talbot mixing, with a spatial phase offset due to the delayed correction from the servo-lag. The corresponding asymmetry is highlighted with red dots in the image plane which is really the inverse Fourier Transform, taken in the Fraunhofer diffraction limit. The large central speckle is due to the constant amplitude, on the left subtracted off from the displayed amplitude so that the small errors are more visible.}
\end{figure}

Now that we have described how free-space propagation affects the complex illumination of the electric field in the Fresnel regime, we will explore an analytic derivation of how this results in an asymmetric PSF for a single mode of phase aberration in one-dimension.

When the inverse of the spatial frequency of the mode is much longer than the wavelength of the propagating light (or $\lambda^2f^2 <\/< 1$), a binomial expansion of the free space propagation transfer function
\begin{equation}
    H(z) \approx e^{-i \pi \lambda f^2 z}
\end{equation}
allows us to perform the analysis up to a constant phase term. In the example of a single mode of sinusoidal phase aberration only (no amplitude errors), with period $p = 1/f$, and a phase variation amplitude $\alpha$, the complex illumination can be written $U = Ae^{i\phi}$ with
\begin{align}
A &= 1 \\
\phi &= \alpha \sin(2\pi x/p) .
\end{align}
It can be shown \cite{Zhou10} that for small phase errors ($\alpha <\/< 1$) that the illumination vector at propagation distance $z$ takes the form $U(z) = A(z) e^{i\phi(z)}$ where the phase and amplitude have appropriately been "mixed" due to the scintillation effects
\begin{align}
A(z) &= 1 + \alpha\sin(2\pi z/z_T)\sin(2\pi x/p) \\
\phi(z) &= \alpha\cos(2\pi z/z_T)\sin(2\pi x/p) .
\end{align}
Here $z_T$ is the Talbot Length
\begin{equation}
    z_T = \frac{\lambda}{1 - \sqrt{1 - \lambda^2f^2}} \approx \frac{2}{\lambda f^2} .
\end{equation}

With the assumption that an ideal AO system corrects phase aberrations after a short delay $t$ due to the servo-lag error, our AO-corrected phase will be the difference between two propagated modes, one shifted along the direction of the wind with the coordinate transform $x \rightarrow x - vt$ if the wind velocity $v$ points along the positive x-axis, and the other the initially measured phase
\begin{eqnarray}
\phi_{\textrm{AO}} &=& \phi_z(x-vt) - \phi_z(x) \\
 &=& \alpha\cos(2\pi z/z_T)\sqrt{2}\sqrt{1-\cos(2\pi vt/p)}\sin\Big(2\pi x/p + \arctan\Big[\frac{-\sin(2\pi vt/p)}{\cos(2\pi vt/p) - 1}\Big]\Big) .
\end{eqnarray}
To simplify this expression, we will use the substitutions
\begin{eqnarray}
\mathcal{A} &=& \alpha\sin(2\pi z/z_T) \\
P &=& \alpha\cos(2\pi z/z_T)\sqrt{2}\sqrt{1-\cos(2\pi vt/p)} \\
\Delta &=& -\frac{p}{2\pi}\arctan\Big[\frac{-\sin(2\pi vt/p)}{\cos(2\pi vt/p) - 1}\Big]
\end{eqnarray}
where $\mathcal{A}$ represents the term that modulates the amplitude, $P$ is the term that modulates the phase, and $\Delta$ is a term that represents a phase shift between the amplitude term and the phase term. With these substitutions, the expressions for our AO-corrected amplitude and phase become
\begin{eqnarray}
A_{\textrm{AO}} &=& 1 + \mathcal{A}\sin(2\pi(x-vt)/p) \\
\phi_{\textrm{AO}} &=& P\sin(2\pi(x-\Delta)/p).
\end{eqnarray}
An example of a single mode of AO-corrected illumination and its corresponding electric field distribution in the image plane is demonstrated in Figure 1. For this figure, the principal observation is how the delayed AO correction and scintillation produce residual errors which are spatially offset, and how the corresponding speckles in the image plane are asymmetric.

\subsection{Open Loop Model Validation}
\begin{figure}[b!]
    \centering
    \includegraphics[width=.65\textwidth]{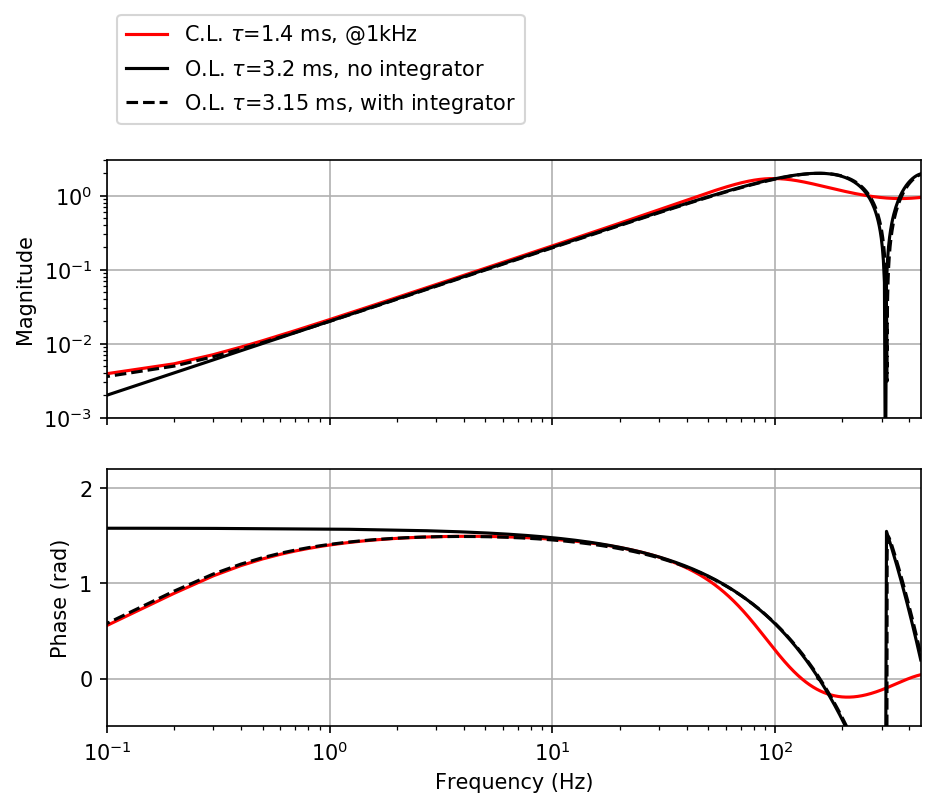}
    \caption{Bode plot for comparing the transfer function of various open and closed loop AO models. By adding an integrating factor to the open-loop model, it becomes possible to recreate the performance of a closed loop system on static errors, although these are outside of the temporal frequency domain we care about, which is approximately from 2-200 Hz. In this range the standard open loop model performs roughly as well as the integrator in mimicking the closed loop model, although the behavior diverges when the inverse of the temporal frequency becomes equal to the time delay.}
\end{figure}
To validate the assumption that an open-loop AO model can accurately reproduce the behavior of a true closed-loop AO system, the frequency responses of both open loop and closed loop AO systems were modeled. Building on the analysis done for GPI's AO system (see section 4.D of \cite{Poyneer16} for a detailed treatment), the standard AO component models and control parameters were used to generate error transfer functions (ETFs). The GPI AO ETF is shown in red in Figure 2. The system has a read-compute delay of 1.4 ms and a maximum controller gain of 0.3. An equivalent open-loop model, black curve in figure, was fit by adjusting the effective delay such the measurement is applied 3.2 ms after it was taken.  As shown in the figure, these two models agree very well in terms of both magnitude and phase response in the range of 2 to 200 Hz.  Given a maximum spatial frequency in the AO-corrected dark hole of $2.78$ m$^{-1}$, these valid temporal frequencies in our model correspond to wind velocities in the range .8 m/s to 72 m/s. These velocities encompass the range of possible wind velocities we might see naturally occurring in the jet stream, which typically ranges from 10 m/s to 60 m/s. The open loop model disagrees at the lowest temporal frequencies, which corresponds to static errors in the system. This discrepancy can be addressed by slightly scaling the phase measurement, e.g. 
\begin{equation}
    \phi_{\textrm{AO}}(t) = \phi(t) - 0.997\phi(t - \tau),
\end{equation}
where $\tau$ is the delay time. The open loop model does not agree at higher temporal frequencies, most obviously
when temporal frequency is the inverse of $\tau$. This is when an ideal open loop AO system coincidentally achieves perfect correction of the translating Fourier mode, which is not physically realizable. This region is beyond our wind speeds of interest, so the open-loop approximation is suitable for our investigation.

\subsection{Fraunhofer Diffraction Limit}
According to Hecht \cite{Hecht2002}, the relationship between the aperture and image planes taken in the far field limit or the Fraunhofer Diffraction limit is simply the Fourier transform, and their derivation results in the expression
\begin{equation}
    E(k_x, k_y) = \iint \limits_{-\infty}^{+\infty} \mathscr{A}(x,y)\textrm{exp}[i(k_x x + k_y y)]\textrm{d}x\textrm{d}y .
\end{equation}
However, Goodman \cite{goodman1996} also claims that the relationship between the two is the Fourier transform, yet they obtain the expression
\begin{equation}
    U(x,y) = \frac{e^{ikz}e^{ik/2z(x^2+y^2)}}{i\lambda z} \iint_{-\infty}^{\infty} U(\xi,\eta) \exp \Big[ -i\frac{2\pi}{\lambda z}(x\xi + y\eta)\Big]\textrm{d}\xi\textrm{d}\eta.
\end{equation}
Comparing the two different expressions, it is not immediately obvious that they are nearly identical. However, after identifying the electric field $E$ and $U$, identifying the geometric relationship between the coordinates in the aperture $(x,y) \rightarrow (\xi,\eta)$ and in the image plane $(k_x, k_y) \rightarrow (\frac{2\pi x}{\lambda z}, \frac{2\pi y}{\lambda z}$), and ignoring the phase prefactors (which do not affect the final intensity), the two answers are comparable with the exception of the sign in the exponent.

Since the choice of defining which Fourier transform is the forward and which is the inverse is arbitrary, both authors choose opposite sign conventions to arrive at the same conclusion that the relationship is the forward transform. However, the physical relationship between the two planes should not have this sign ambiguity. This difference is traceable to an assumption at the beginning of the derivations, where the choice of the direction of phasor rotation in a spherically converging or diverging wave $e^{\pm i(kr-\omega t)}/r$ is used as a Green's function to solve the Huygens-Fresnel diffraction integral.

In practice, the difference between the forward and inverse transforms is essentially a coordinate transform from $x \rightarrow -x$ and $y \rightarrow -y$, and so the orientation of the PSF is flipped along both axes. For most cases, where the PSF is symmetric, this does not matter. However, for our purposes, properly orienting the PSF is critical and so taking note of this fact is important. Later, we will show that in order to remain consistent with observations, the Fraunhofer diffraction limit should use the inverse transform, with a positive sign in the exponent, which results in the stronger asymmetric lobe of the PSF on the opposite side as the wind.

\subsection{Taylor Expansion of the Single Mode PSF}
\begin{figure}[b!]
    \centering
    \includegraphics[width=\textwidth]{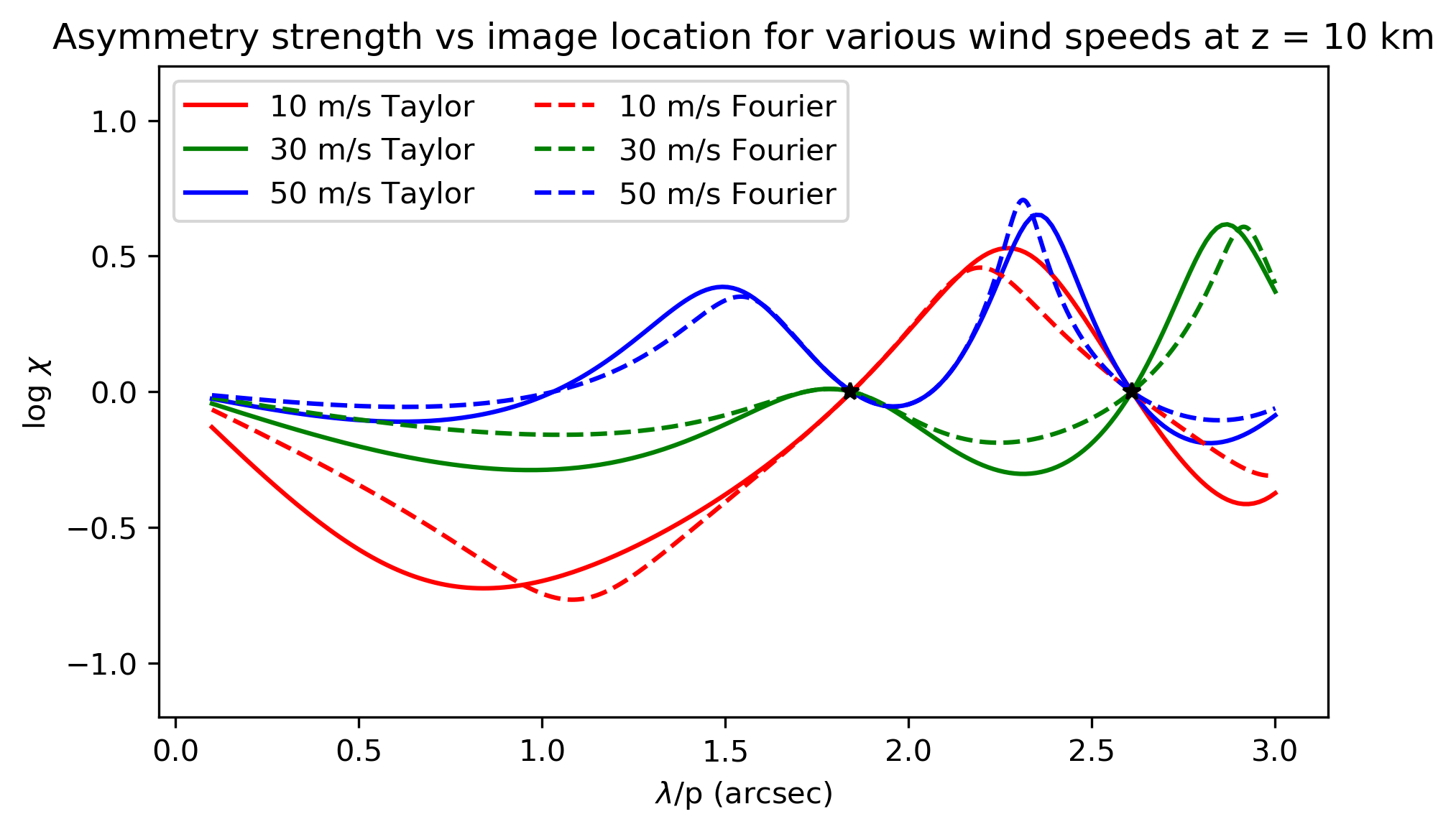}
    \caption{The log of the speckle asymmetry ratio as a function of PSF location for parameters $\alpha = .01$ rad, $t = 3.2$ ms, and $z = 10$ km. The solid line represents the second-order  Taylor expansion and the dashed line represents the numerical solution found using the discrete Fourier Transform. The x-axis is a proxy for mode length, transformed in PSF location when imaged in H-band at $\lambda = 1.6 \mu$m. The zeros corresponding to the particular mode lengths $p$ for which the propagation distance $z$ is an integer multiple of $z_T/4$ are plotted in black stars. For these, PSF$_1 = 0$ because either $\mathcal{A} \sim \sin(2\pi z/z_T) = 0$ or $P \sim \cos(2\pi z/z_T) = 0$, resulting in a symmetric PSF. Although there are additional zeros when the velocity times time delay is comparable to the mode length, visible in both the $30$ m/s and $50$ m/s case, though their analytic solution is more complicated (set $\frac{2\pi}{p}(\Delta - vt) = \pi/2 + n\pi, n \in \mathbb{Z}$).}
\end{figure}

While it may be possible to calculate the AO-corrected electric field distribution in the image plane by taking the inverse Fourier transform of the AO-corrected aperture illumination  by expanding it as an infinite series of Bessel functions, it is not clear that this expression can be easily squared to get the intensity distribution due to an infinite amount of cross-terms. Instead we will follow the conventions of Sivaramakrishnan et al. \cite{Sivaramakrishnan2002} and Perrin et al. \cite{Perrin2003} to arrive at the intensity. However, both Sivaramakrishan et al. and Perrin et al. assume that the image plane PSF is the Fourier transform of the aperture plane, and not the inverse. For all of their results, this fact does not matter but for ours we must be cautious, and remember to apply the appropriate coordinate transform to recover the proper result.
For small $\phi$, the intensity distribution or PSF $= |\mathcal{F}[u]|^2$  can be expanded in a Taylor series 
\begin{eqnarray}
\textrm{PSF} &\approx& \textrm{PSF}_0 + \textrm{PSF}_1 + \textrm{PSF}_{2,\textrm{halo}} + \textrm{PSF}_{2,\textrm{strehl}},
\end{eqnarray}
whose first few terms are
\begin{eqnarray}
\textrm{PSF}_0 &=& aa^* \\
\textrm{PSF}_1 &=& 2\textrm{Im}[a(a^* \star \Phi^*)] \\
\textrm{PSF}_{2,\textrm{halo}} &=& (a \star \Phi)(a^* \star \Phi^*) \\
\textrm{PSF}_{2,\textrm{strehl}} &=& -\frac{1}{2}[a(a^* \star \Phi^* \star \Phi^*) + a^*(a \star \Phi \star \Phi)] .
\end{eqnarray}
Here the case change is used as shorthand for the Fourier transform, so $a = \mathcal{F}[A]$ and $\Phi = \mathcal{F}[\phi]$. Additionally, $\star$ is the convolution operation and $*$ is the complex conjugate. For our AO-corrected illumination, these can be calculated as follows
\begin{equation}
a_{\textrm{AO}} = \delta(f) +\frac{\mathcal{A}}{2i}\delta(f-1/p)e^{-i\frac{2\pi}{p}vt}-\frac{\mathcal{A}}{2i}\delta(f+1/p)e^{i\frac{2\pi}{p}vt}
\end{equation}
\begin{equation}
\Phi_{\textrm{AO}} = \frac{P}{2i}\delta(f-1/p)e^{-i\frac{2\pi}{p}\Delta}-\frac{P}{2i}\delta(f+1/p)e^{i\frac{2\pi}{p}\Delta},
\end{equation}
where $\delta$ is the Dirac delta distribution. It is worthwhile to note that here we implicitly are using the entire real line in the Fourier transform, which can be interpreted as using an infinitely large telescope, or as a telescope with an idealized perfect coronagraph. This leads to a PSF with terms
\begin{flalign}
& \textrm{PSF}_0 = \delta(f) + \frac{\mathcal{A}^2}{4}\Big[\delta(f-1/p)+\delta(f+1/p)\Big] &\\
& \textrm{PSF}_1 =  \frac{-\mathcal{A}P}{2}\sin\Big(\frac{2\pi}{p}(vt-\Delta)\Big)\Big[\delta(f-1/p)-\delta(f+1/p) \Big] &\\
& \textrm{PSF}_{2,\textrm{halo}} = \frac{P^2}{4}\Big[δ(f+1/p) + δ(f-1/p)\Big]  + \frac{\A^2 P^2}{16}\Big[2\Big(1 + \cos\Big(\frac{4π}{p}(vt -Δ)\Big)\Big)δ(f)
    + δ(f + 2/p) + δ(f - 2/p)\Big]  &\\
& \textrm{PSF}_{2,\textrm{strehl}} = \frac{-P^2}{2}\delta(f) -\frac{\mathcal{A}^2P^2}{32}\Big(4 + 2\cos\Big(\frac{4\pi}{p}(vt-\Delta)\Big)\Big)\Big[\delta(f-1/p)+\delta(f+1/p)\Big].&
\end{flalign}
Indeed the even order terms are symmetric and the first odd order term PSF$_1$ is responsible for the observed asymmetry. We will examine the ratio of the amplitude for the speckles on either side of the image, and to do so evaluate the PSF at the location appropriate $f=\pm1/p$
\begin{align}
\textrm{PSF}_0(f=-1/p) = \textrm{PSF}_0(f=1/p) &= \frac{\mathcal{A}^2}{4} \\
- \textrm{PSF}_1(f=-1/p) = \textrm{PSF}_1(f=1/p) &=  \frac{-\mathcal{A}P}{2}\sin\Big(\frac{2\pi}{p}(vt-\Delta)\Big)\\
\textrm{PSF}_2(f=-1/p) = \textrm{PSF}_2(f=1/p) &= \frac{P^2}{4} - \frac{\mathcal{A}^2P^2}{32}\Big[4 + 2\cos\Big(\frac{4\pi}{p}(vt-\Delta)\Big)\Big] .
\end{align}
Our metric for the ratio of the right to left speckle asymmetry is the Taylor expansion sum of the PSF terms evaluated at these appropriate locations
\begin{equation}
\chi = \frac{\textrm{PSF}(f=1/p)}{\textrm{PSF}(f=-1/p)} ,
\end{equation}
which is a function of propagation distance $z$, mode period $p$, and velocity times delay $vt$. A comparison of the asymmetry metric for both the numerical single mode scintillation using the discrete Fourier transform and our analytic second order Taylor expansion is given in Figure 3.

Highlighting a few observations from this plot: first, we note that for wind layers at $z = 10$ km, the first zero crossing is not until roughly 1 arcsec in the image, and so for most of the relevant observations, the strongest asymmetry will be on the side opposite to the direction of the wind, though for higher spatial frequencies corresponding to the edges of the PSF in the image plane, this will not always be true. Second, the strength of the asymmetry is greater for slower wind velocities in this region of interest. Third, it is worthwhile to note that our Taylor expansion generally gets the behavior of the asymmetry analytically correct, although differs from the numerical solution due to the absence of higher order terms in the expansion.

\section{Atmospheric Turbulence Scintillation Simulation}

The response of an ideal open loop AO system to frozen flow Kolmogorov turbulence was developed in simulations using the method of the angular spectrum free space propagation and Fourier decomposition. These simulations can place many layers of Turbulence at arbitrary altitudes, and we explore the effects of single and multiple layers. Readers interested in many of the precise details of these simulations are referred to Appendix A, which discusses our techniques at greater length. However, here we will discuss the results of these simulations in the context of the previous analysis.

To start, the effect of scintillation in the halo of an AO PSF is most clearly seen for the case of a single layer
\begin{figure}[h!]
    \centering
    \includegraphics[width=\textwidth]{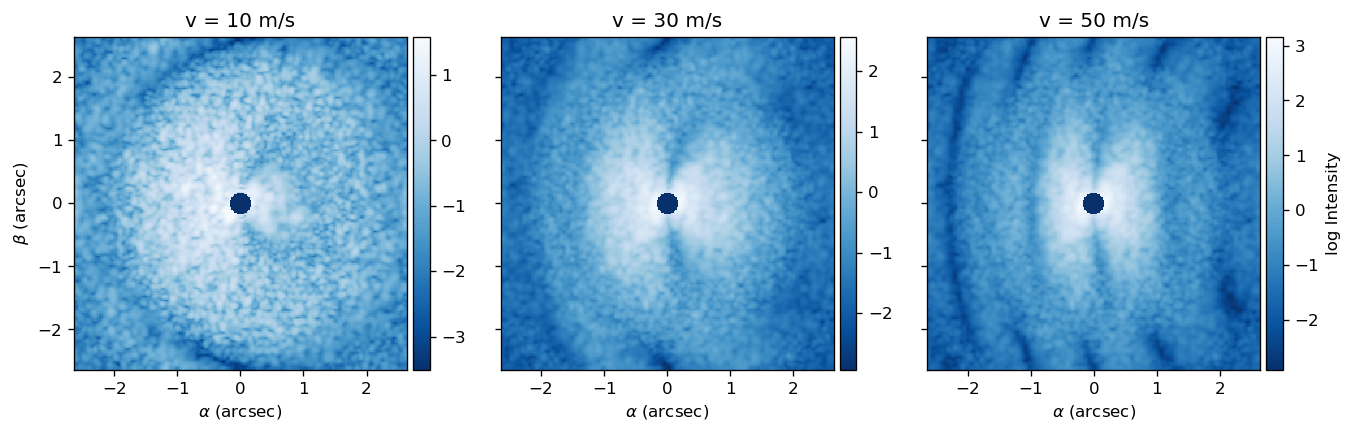}
    \caption{For a single layer of atmospheric turbulence at $z = 10$ km, the scintillation halo at low wind velocities is highly apparent, and still noticeable at rapid velocities. Each image has its own unique colorbar, so that the variation in the halo shape is visible, although from looking at the magnitude of the halo intensity, it is clear that slower wind velocities are corrected better in the metric of total scattered light.}
\end{figure}
\begin{figure}[h!]
    \centering
    \includegraphics[width=\textwidth]{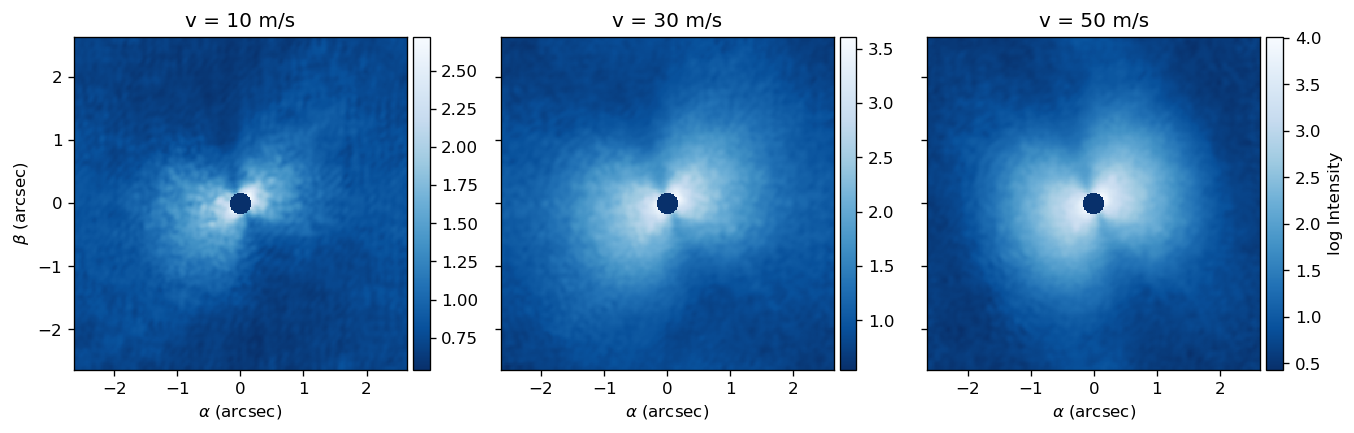}
    \caption{For simulations with many wind layers, here L = 18 is the number of layers used, the effect of the scintillation asymmetry is significantly less apparent, although at low wind velocities where the asymmetry is strongest it is still visible. This is likely due to interference from the other wind layers causing the delicate spatial offset in the illumination to become washed out when combined with the finite sampling resolution of our simulation, and for real observations, the detector.}
\end{figure}
of Kolmogorov turbulence, which is plotted in Figure 4 for three characteristic wind velocities. The single layer simulation uses a turbulent layer altitude of $z = 10$ km, which roughly corresponds to the altitude of the jet stream. The layer altitude equals the propagation distance of the light when the telescope is pointed at zenith, however, for low elevation pointing, the propagation distance and therefore strength of the scintillation, will grow larger. The jet stream is the strongest layer of turbulence which has enough relevant altitude to scintillate significantly (recall the strength of the scintillation amplitude errors grow like $\sin(2\pi z/z_T)$).

The wind in the single layer example is propagating directly to the right, along the positive x-axis. For the multi-layer case, the wind velocity for the jet stream is generally pointing towards 2 o'clock. As a consequence, we can see that the brighter asymmetric lobe of the PSF is on the opposite side of wind direction. To remain consistent with observations, it is necessary that the image plane is the inverse Fourier transform of the aperture plane in the Fraunhofer limit. Looking at just the single layer, it is apparent that the scintillation dominates for slower wind velocities. This scintillation halo would be close to the true PSF if the atmosphere was actually only a single layer of turbulence. Examining the edges of the PSF, one can identify the region where the asymmetry metric $\log \chi = 0$ that was described analytically, noticeable here as dark bands.

The halos for the single layer case do not match real observations well. However, when we run this simulation with many wind layers, each with unique velocities pulled from an instance of the NOAA GFS to mimic real observing conditions, the extreme scintillation halo washes out. This is likely an interference between multiple independent Kolmogorov layers masking the delicate spatial offset for a single mode needed to create the asymmetry demonstrated earlier. While at quick glance the PSF does not appear to vary with the scaled wind velocities cited in their titles, upon closer inspection the asymmetry remains, and is most obvious for the slowest wind velocity. For most observations, with jet stream velocities upwards of $50$ m/s, there are no apparent deviations from the symmetry of the butterfly shaped halo that is often seen. But in the rare cases when the high-altitude winds are very slow, around $10$ m/s, the scintillation becomes stronger and introduces a noticeable asymmetry in the PSF along the axis of the wind propagation.

\section{Observational Correlations}
\begin{figure}[h!]
    \centering
    \includegraphics[width=\textwidth]{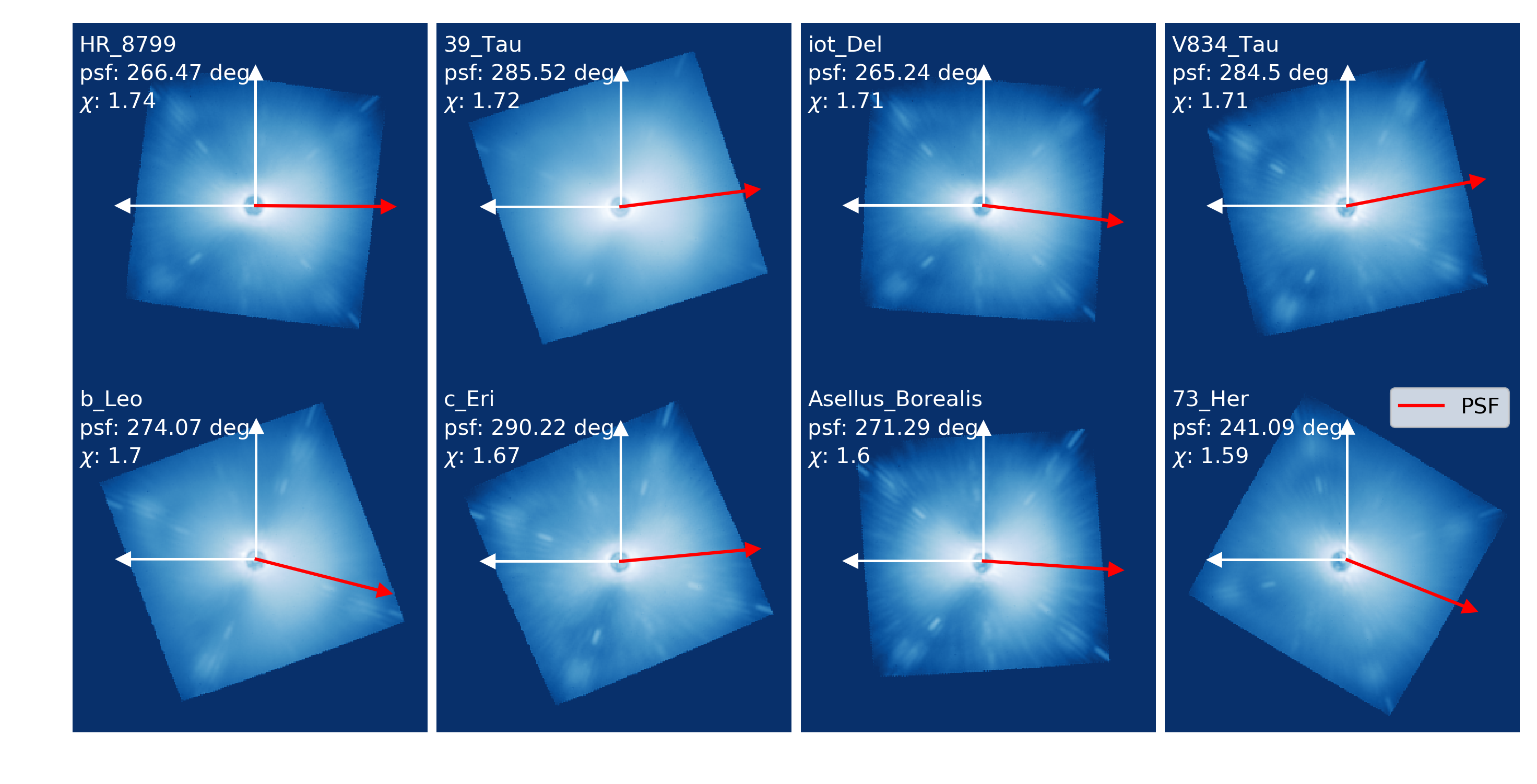}
    \caption{A selection of images taken with the Gemini Planet Imager, ordered by decreasing asymmetry metric $\chi$. Each image is the most strongly asymmetric image taken from each set of observations of a single target, such that there are no duplicates, which shows that this effect is often recurring, and not limited to single cases. In addition, each image has been rotated and flipped such that North is up and East is to the left. The angular size of each GPI image is 2.8 arsec $\times$ 2.8 arcsec. In each image the direction of the strong asymmetry is plotted with a red arrow, and the corresponding direction in degrees azimuth projected onto the ground is printed in the corner of each image, alongside the ratio of the asymmetry metric $\chi$. Glancing over the entire dataset, it is readily apparent that the strong asymmetry direction is predominately pointing into the West.}
\end{figure}
\begin{figure}[h!]
    \centering
    \includegraphics[width=.6\textwidth]{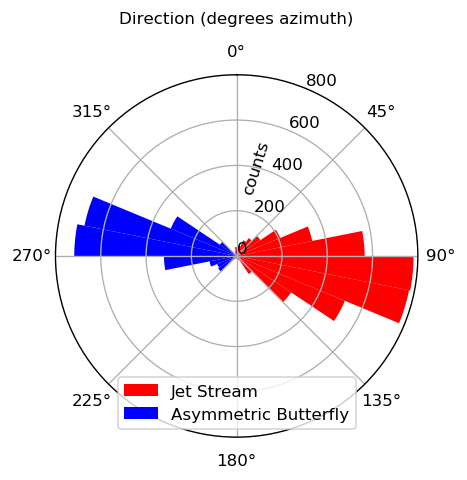}
    \caption{Angular histogram of the asymmetric butterfly vector's projection onto the ground for our sample in degrees from azimuth alongside the direction of the jet stream. The asymmetric butterfly nearly always points West while the jet stream is nearly always points East, which is only possible if the image plane electric field distribution is the inverse Fourier transform of the aperture plane. These distributions are not temporally matched but are rather the entire subset observations with strong asymmetry and the entire distribution of jet stream wind directions over the course of the survey. For temporally matched correlations, see Figures A2 and 8.}
\end{figure}
\begin{figure}[h!]
    \centering
    \includegraphics[width=.8\textwidth]{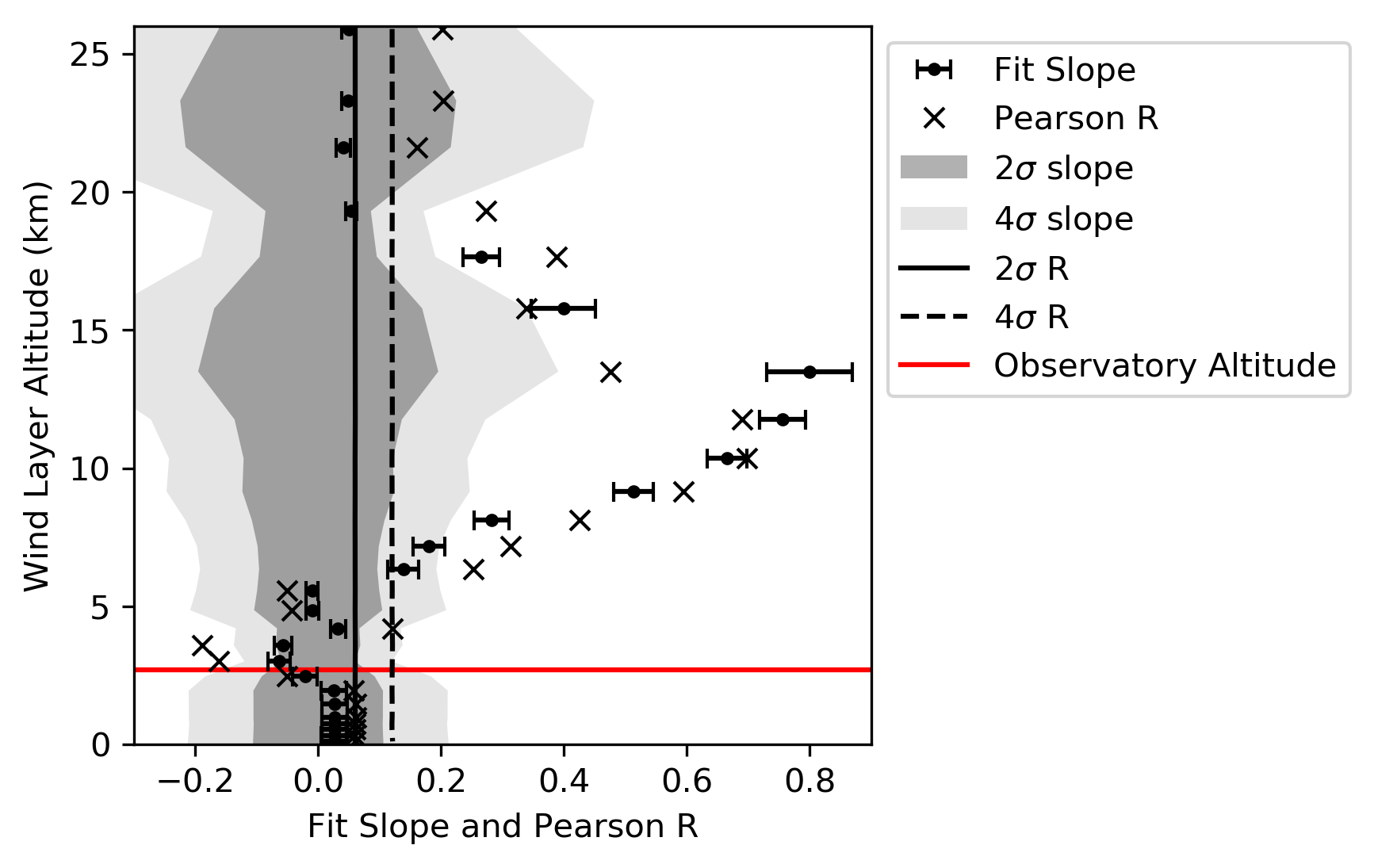}
    \caption{Distilling the information from the scatter plots in Figure A2 into their slopes and Pearson R coefficient as a function of altitude from the GFS. The model contains layers all the way to sea level despite the observatory being around 3 km up because of the uniform grid spacing. The strong correlations are visible for the altitudes relevant to the jet stream in both the slope of the best fit line and the R coefficient. These correlations are much greater than 2-4 sigma chances concerning the null hypothesis, generated from bootstrap sampling at random, shown as the shaded regions and solid lines for the slope and the R coefficient, respectively.}
\end{figure}
\begin{figure}[h!]
    \centering
    \includegraphics[width=\textwidth]{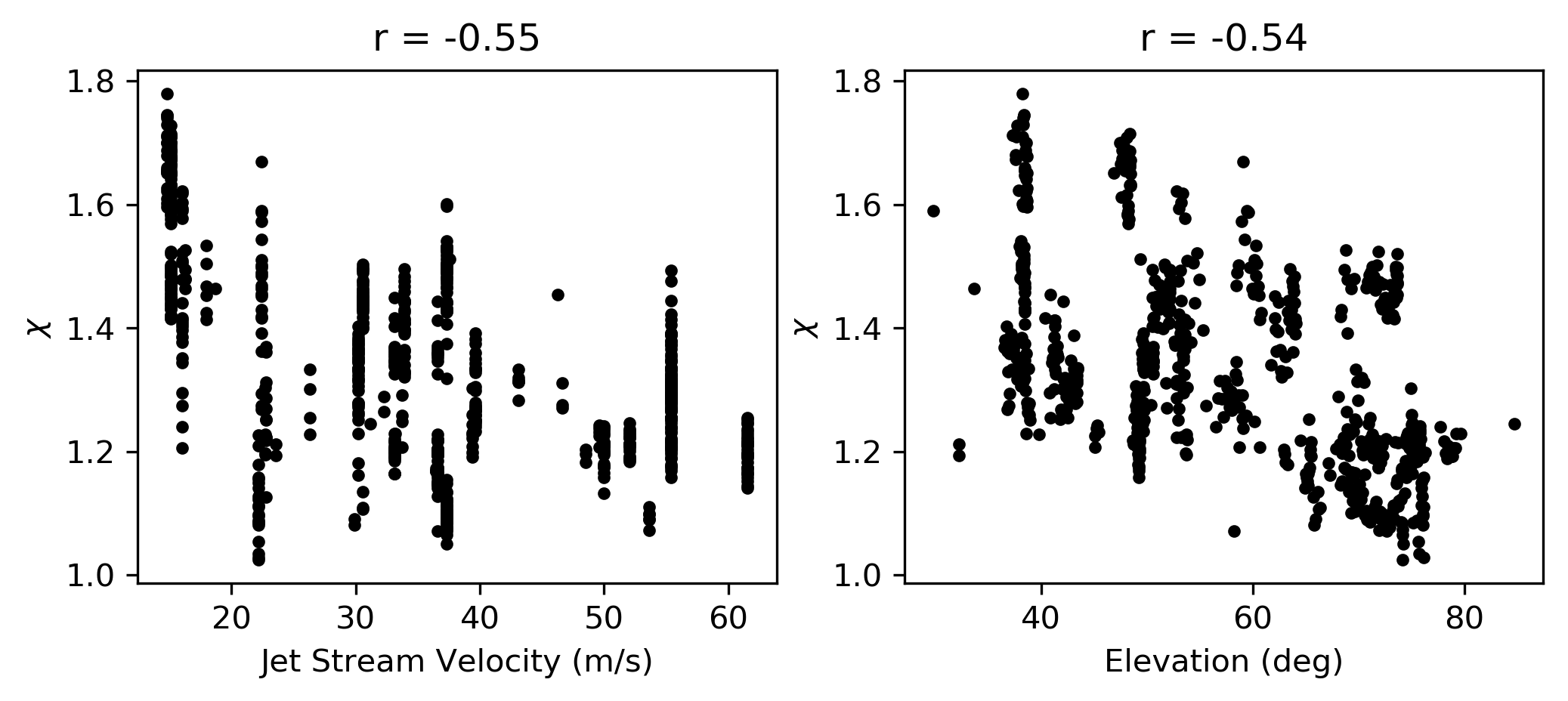}
    \caption{Asymmetry strength $\chi$ when compared to the velocity of the jet stream and the telescope elevation for our sample. Both exhibit strong correlations which corroborate our analytic understanding of the origin of the asymmetry. Slower jet stream velocities directly cause stronger asymmetry, while decreased telescope elevation has two effects. One to decrease the apparent wind velocity and the other to push the turbulent layers father away, giving more distance to scintillate.}
\end{figure}

The previously discussed asymmetry for AO PSFs is observable in real data taken from observations using the Gemini Planet Imager. Over the course of a few years of observations, over 20,000 images were taken in H-band as a part of the GPI Exoplanet Survey. We exercise selection cuts to find a sample of PSFs exhibiting this characteristic asymmetry using the following methods.

The first metric is the fractional standard deviation (FSD) in an annulus 30 to 70 pixels wide centered around the star's location. We first select images with FSD greater than one standard deviation above the mean. This selects for images where the relatively fast jet stream turbulence was dominant, producing PSFs with the characteristic butterfly shaped halo. Then, we construct a angular profile of the image by integrating along the radial axis. This allows us to fit for the preferential wind direction modulo 180 degrees. These techniques follow exactly from our previous work \cite{Madurowicz2018}, yet in this work we expand by defining the asymmetry metric $\chi$ between the two lobes of the PSF. $\chi$ is defined as the ratio of the summed intensities in the half-annuli perpendicular to the axis of wind propagation. This asymmetry metric is used to break the 180 degree symmetry of the wind axis to find the true asymmetric butterfly vector in the image plane. In addition, our metric $\chi$ is folded around 1 by inversion such that the half-annulus with greater intensity is always in the numerator, meaning $\chi$ always takes on values greater than 1, and larger values correspond to greater asymmetry. Note that this $\chi$ is not identical to the $\chi$ described earlier for the one dimensional case, although both are intensity ratios, this one lives in two dimensions. A large sample of PSFs selected in this manner as displayed in Figure 6.

These asymmetric butterfly vectors are projected from the image plane onto the surface of the Earth, for comparison with meteorological wind data. Readers interested in the trigonometric problem of relating the two are referenced to our techniques in Appendix B. The directions and velocities of the wind for various layers in the Earth's atmosphere are taken from the NOAA Global Forecast System \cite{GFS}, archives of which are available to the public. The distribution of wind directions for the jet stream and the distribution of directions for the asymmetric butterfly vector projected on the ground are displayed in Figure 7 for a concise comparison. It is apparent that the jet stream predominately points East in accordance with its origin due to the rotation of the Earth. As a consequence the resulting asymmetric butterfly vector points entirely towards the West. This observation confirms our prior analysis regarding the strength of the asymmetry being opposed to the direction of the wind.

To further verify this correlation between the jet stream and the asymmetric butterfly, we take our sample of data where $\chi \geq 1.2$, which is where the asymmetry begins to be noticeable to the eye, and match the time of observation to time of the wind data for the approximate location of Cerro Pachon, Chile, where the Gemini South Telescope takes it observations from, and plot the resulting correlations for the various directions of the wind layers in Figure A2. For most of the wind layers in the NOAA GFS, there is little to no correlation between the direction of the asymmetric butterfly and the direction of the wind, with the exception of the wind layers around 100-300 hPa, which roughly corresponds to altitudes of 10-15 km, which is the approximate altitude of the jet stream. The strength of these correlations, as measured by the Pearson R coefficient, and the slopes of the best fit lines are plotted for a concise summary in Figure 8. The rise of the correlations well beyond the limits imposed by a null hypothesis bootstrap show that these strong correlations are not spurious, and a very real phenomenon.

To further explore our observations of the asymmetry, we plot $\chi$ versus the velocity of the jet stream and the elevation of the telescope pointing for our subset of observations which have been temporally matched to the wind data in Figure 9. Both plots have reasonable correlation strength with R coefficient around negative one half, and both of these correlations have an intuitive sense. Our analysis showed that the asymmetry should be strong for slower wind velocities, and this is indeed verified through the first plot. The second has contributions from two different effects. When the telescope is pointing towards zenith (elevation = 90 deg), the turbulent wind layers in the jet stream are as close to the telescope as possible along the line of sight. As a consequence, the propagation distance $z$ will be smaller relative to the Talbot length $z_T$ in comparison to when the telescope is pointed lower towards the horizon. A lower elevation value corresponds to the telescope effectively pushing the layer for a given altitude to a further propagation distance, giving the light more time to scintillate. This pushes the relevant mode lengths for the asymmetry closer to the center of the image, see Figure A3 for a remake of Figure 3 with z = $25$ km. In addition, a low telescope elevation also has the effect of changing the apparent wind velocity over the aperture. A telescope pointing away from zenith can only  observe relatively slower wind velocities than one pointing directly up as long as the winds are parallel to the surface of the Earth. These two effects work together to build the strong correlation between telescope elevation and asymmetry.

\newpage
\section{Discussion}

High contrast imaging systems, with large actuator counts, are often limited by time lag errors. In particular, when observing bright stars, they are the dominant source of scattered light within the “dark hole” region \cite{Bailey2016}. Under mid-latitude Chilean conditions, the velocity of the jet stream is often the dominant source of these errors, even if its contribution to the total $r_0$ is moderate. This implies different observatory sites with slower winds may have a comparative advantage, as well as the merits in scheduling observations around poor atmospheric conditions. In our previous work \cite{Madurowicz2018}, we demonstrate that the jet stream is highly correlated with these errors, in a very large sample of observations.

Scintillation has previously been identified as a performance limiting factor in high contrast imaging \cite{guyon2005}, and here we have demonstrated severity of this effect, visible in the form of PSF asymmetry. This has implications for both current and future adaptive optics systems, especially but not exclusively those designed for high-contrast imaging. Many analyses often assume that various sources of scattered light are uncorrelated \cite{guyon2005} for simplicity. In this paper, we show that correlations between scintillation amplitude errors and time lag phase errors exist and can dominate during ideal conditions. If one simply tries to minimize the time lag error alone, without additionally compensating amplitude errors, the asymmetry will become larger as the effective wind velocity is decreased. This effect is worst when considering the asymmetry in low order modes which correspond to small separations in the final images, the region where planets or protoplanetary disks are most likely to be found. While it is still optimal to have as-fast-as-possible correction in the metric of total scattered light in the halo, as computers get faster and algorithms are optimized the asymmetry will begin to play a larger role relative to other errors in the final PSF formed in adaptive optics images, leading many to explore possible routes for correction.

Scintillation errors are uncorrectable for an AO system operating in single-DM phase-conjugation mode, even if it is infinitely fast (see Figure A4). This in turn will set a performance floor for such systems, and the asymmetry detected here provides a first measurement of the level at which those effects begin to dominate. One could address this with a system that corrects amplitude errors as well, using the Talbot or scintillation mixing effect to our advantage. Having two deformable mirrors at two unique conjugate planes in the optical system enables some phase introduced by one DM to transform into amplitude, allowing one to correct amplitude errors from scintillation. This concept has been proposed for space-based coronagraphs \cite{Jonge2018} to correct static amplitude errors, and preliminary laboratory testing is underway \cite{Seo2018}. Similar designs have been expressed for improvements in laser communications \cite{Wu18}\cite{roggemann1998}. But we are particularly interested in the future of high-contrast imaging, particularly in the era of ELTs. Such a system could also correct static amplitude errors in an ELT, such as reflectivity variations between segments which have been re-coated at different times \cite{macintosh2006}\cite{Troy2006}. Driving such a system would require knowledge of both the phase and amplitude of the science wavefront.  Space-based coronagraphic instruments can achieve this using the science camera and making several measurements while modulating the speckle field with the DM \cite{Borde_2006}\cite{Giveon09}. Another similar focal plane wavefront sensing approach was recently proposed on ground-based telescopes \cite{Gerard2018a}, which could also correct for amplitude aberrations from scintillation. Conventional Shack-Hartmann wavefront sensors measure some intensity information but this is complicated by spots in each subaperture moving out the active pixels of the detector, so it is impractical to separate phase and amplitude.  Other methods to estimate the amplitude errors could utilize fast interferometric focal-plane sensing \cite{Gerard2018b}, or something as simple as adding a high-speed direct pupil-imaging channel to a traditional adaptive optics wavefront sensor. 

The scintillation halo observed in AO PSFs is a challenge for effective post-processing of datasets. Many algorithms are designed to subtract a static speckle field with respect to the detector plane whose origin is from optical phase errors from imperfections, misalignment, non-common-path errors (NCPE), and other systematic sources. Since these speckles have a unique spectral dependence, moving to farther separations at longer wavelengths, instruments imaging with an integral field spectrograph (IFS) can measure that spectral dependence and remove those aberrations. However, since the scintillation halo intensity is driven by the effectiveness of the AO system, the halo appears fainter at longer wavelengths. Because the halo originates from a spectrum of turbulent modes which decide their final image plane location, the halo does not scale in image location with wavelength, unlike the static speckles best removed with SDI \cite{Marois2006b}. In addition, the scintillation halo is fixed with respect to the direction of the high altitude winds, and it does not track with the rotation of the instrumental errors or the parallactic rotation of the astrophysical signal in an observing strategy such as ADI. \cite{Marois2006a}

Various different post processing algorithms \cite{pyklip}\cite{Pueyo2012} often use a high pass filter (HPF) to attempt to eliminate the diffuse background halo, and this is effective for regions of the image at large separations. However, near the coronagraphic mask, the scintillation can have very sharp features, demonstrated analytically as regions where $\log \chi = 0$, and observable in simulations as dark regions perpendicular to the axis of the wind direction. When an HPF that preserves the features of a planet is applied to this halo, residuals which vary on spatial scales comparable to the planet are not removed. Often, a quadrupolar residual artifact near the coronagraphic mask if left which is large compared to the speckle residuals in the smooth halo at large separations. These residuals contribute significant noise to planetary detection attempts at the nearest separations, where the likelihood of detection is highest from their population distributions. \cite{Nielsen19} In addition, imaging extended objects such as debris disks cannot utilize a HPF in post-processing, implying a limit to sensitivity even at wide separations for diffuse unpolarized structure.

Various methods of subtracting the scintillation halo have been suggested. One could take an empirical approach, using a PCA style analysis to model the shape of the halo over an averaged population of observations. Acknowledging this effect is unique and must be treated independently with this sort of approach can be effective at improving the final SNR in your detection algorithm. \cite{Nielsen19} Another approach may be to model the PSF end-to-end with complete simulations of the instrument and atmosphere, although this approach is significantly limited by the extent to which your simulated instrument can account for all real sources of error. Not only do errors arise from instrumental effects like DM fitting and NCPE, but also the non-Kolmogorov deviations in the turbulent spectrum from environmental effects \cite{Tallis2018}, as well as the finite temporal resolution of available atmospheric information. Another path may attempt to estimate the PSF using a reconstruction from measured AO telemetry. However, with current WFS measurements, estimating the amplitude error from scintillation is rather difficult, as current instruments are not designed to measure wavefront amplitude. It is likely that the optimal approach to handling these errors is at the instrument level itself, as discussed previously, with a method to measure and correct the wavefront amplitude in real time. As high-contrast imaging strives for higher and higher performance levels, identification, measurement, estimation, and mitigation of scintillation errors will become increasingly important.

\bibliographystyle{plain}
\bibliography{main}

\begin{thebibliography}{10}

\bibitem{Blackman}
R.~B. Blackman and J.~W. Tukey.
\newblock The measurement of power spectra from the point of view of
  communications engineering — part i.
\newblock {\em Bell System Technical Journal}, 37(1):185--282.

\bibitem{Borde_2006}
Pascal~J. Borde and Wesley~A. Traub.
\newblock High-contrast imaging from space: Speckle nulling in a low-aberration
  regime.
\newblock {\em The Astrophysical Journal}, 638(1):488--498, feb 2006.

\bibitem{Cantalloube2018}
F~Cantalloube, Emiel Por, Kjetil Dohlen, Jean-François Sauvage, A~Vigan,
  M~Kasper, N~Bharmal, Th~Henning, W~Brandner, Julien Milli, C~Correia, and
  T~Fusco.
\newblock Origin of the asymmetry of the wind driven halo observed in
  high-contrast images.
\newblock {\em Astronomy \& Astrophysics}, 11 2018.

\bibitem{Cantalloube2018B}
Faustine {Cantalloube}, Marie {Ygouf}, Laurent {Mugnier}, David {Mouillet},
  Olivier {Herscovici-Schiller}, and Wolfgang {Brandner}.
\newblock {Status of the MEDUSAE post-processing method to detect circumstellar
  objects in high-contrast multispectral images}.
\newblock {\em arXiv e-prints}, page arXiv:1812.04312, December 2018.

\bibitem{Jonge2018}
Chris de~Jonge, Pierre Baudoz, Raphaël Galicher, Robert Huisman, Reynier
  Peletier, and Bayu Jayawardhana.
\newblock Effect of multiple deformable mirrors in broadband high-contrast
  coronagraphs.
\newblock volume 10703, pages 10703 -- 10703 -- 10, 2018.

\bibitem{Ruffio2017}
Jean-Baptiste~Ruffio et~al.
\newblock Improving and assessing planet sensitivity of the gpi exoplanet
  survey with a forward model matched filter.
\newblock {\em The Astrophysical Journal}, 842(1):14, 2017.

\bibitem{Nielsen19}
Nielsen et~al.~in review.
\newblock The gemini planet imager exoplanet survey: Giant planet and brown
  dwarf demographics from 10–100 au.

\bibitem{Gerard2018a}
Benjamin~L. {Gerard}, Christian {Marois}, and Rapha{\"e}l {Galicher}.
\newblock {Fast Coherent Differential Imaging on Ground-based Telescopes Using
  the Self-coherent Camera}.
\newblock {\em aj}, 156:106, Sep 2018.

\bibitem{Gerard2018b}
Benjamin~L. Gerard, Christian Marois, Raphaël Galicher, and Jean-Pierre
  Véran.
\newblock Fast focal plane wavefront sensing on ground-based telescopes.
\newblock volume 10703, pages 10703 -- 10703 -- 11, 2018.

\bibitem{Giveon09}
Amir Give'on.
\newblock The electric field conjugation - a unified formalism for wavefront
  correction algorithms.
\newblock In {\em Frontiers in Optics 2009/Laser Science XXV/Fall 2009 OSA
  Optics \& Photonics Technical Digest}, page AOWA3. Optical Society of
  America, 2009.

\bibitem{goodman1996}
J.W. Goodman.
\newblock {\em Introduction to Fourier Optics}.
\newblock McGraw-Hill Series in Electrical and Computer Engineering:
  Communications and Signal Processing. McGraw-Hill, 1996.

\bibitem{guyon2005}
Olivier Guyon.
\newblock Limits of adaptive optics for high-contrast imaging.
\newblock {\em The Astrophysical Journal}, 629(1):592, 2005.

\bibitem{Hardy1998}
J.W. Hardy.
\newblock {\em Adaptive Optics for Astronomical Telescopes}.
\newblock Oxford Series in Optical \& Ima. Oxford University Press, 1998.

\bibitem{Hecht2002}
Eugene Hecht.
\newblock {\em Optics}.
\newblock Addison-Wesley, 2002.

\bibitem{Johansson1994}
Erik~M. Johansson and Donald~T. Gavel.
\newblock Simulation of stellar speckle imaging.
\newblock volume 2200 of {\em Proc. SPIE}, pages 2200 -- 2200 -- 12, 1994.

\bibitem{macintosh2006}
B.~{Macintosh}, M.~{Troy}, R.~{Doyon}, J.~{Graham}, K.~{Baker}, B.~{Bauman},
  C.~{Marois}, D.~{Palmer}, D.~{Phillion}, L.~{Poyneer}, I.~{Crossfield},
  P.~{Dumont}, B.~M. {Levine}, M.~{Shao}, G.~{Serabyn}, C.~{Shelton},
  G.~{Vasisht}, J.~K. {Wallace}, J.-F. {Lavigne}, P.~{Valee}, N.~{Rowlands},
  K.~{Tam}, and D.~{Hackett}.
\newblock {Extreme adaptive optics for the Thirty Meter Telescope}.
\newblock In {\em Society of Photo-Optical Instrumentation Engineers (SPIE)
  Conference Series}, volume 6272 of {\em procspie}, page 62720N, June 2006.

\bibitem{Madurowicz2018}
Alexander Madurowicz, Bruce~A. Macintosh, Jean-Baptiste Ruffio, Jeffery
  Chilcote, Vanessa~P. Bailey, Lisa Poyneer, Eric Nielsen, and Andrew~P.
  Norton.
\newblock Characterization of lemniscate atmospheric aberrations in gemini
  planet imager data.
\newblock volume 10703 of {\em Proc. SPIE}, pages 10703 -- 10703 -- 13, 2018.

\bibitem{MalesGuyon17}
Jared~R. Males and Olivier Guyon.
\newblock Ground-based adaptive optics coronagraphic performance under
  closed-loop predictive control.
\newblock {\em Journal of Astronomical Telescopes, Instruments, and Systems},
  4:4 -- 4 -- 21, 2018.

\bibitem{Marois2006a}
Christian Marois, David Lafreniere, Rene Doyon, Bruce Macintosh, and Daniel
  Nadeau.
\newblock Angular differential imaging: A powerful high-contrast imaging
  technique.
\newblock {\em The Astrophysical Journal}, 641(1):556--564, apr 2006.

\bibitem{Marois2006b}
Christian Marois, Don~W. Phillion, and Bruce Macintosh.
\newblock Exoplanet detection with simultaneous spectral differential imaging:
  effects of out-of-pupil-plane optical aberrations.
\newblock volume 6269, pages 6269 -- 6269 -- 11, 2006.

\bibitem{GFS}
NOAA NCEP.
\newblock Global forecast system analysis dataset.
  \url{ftp://nomads.ncdc.noaa.gov/GFS/analysis\_only/}.

\bibitem{Bailey2016}
Vanessa P.~Bailey, Lisa A.~Poyneer, Bruce A.~Macintosh, Dmitry Savransky, Jason
  Wang, Robert J.~De~Rosa, Katherine B.~Follette, S~Mark~Ammons, Thomas
  Hayward, Patrick Ingraham, Jérôme Maire, David W.~Palmer, Marshall
  D.~Perrin, Abhijith Rajan, Fredrik Rantakyro, Sandrine Thomas, and
  Jean-Pierre Véran.
\newblock Status and performance of the gemini planet imager adaptive optics
  system.
\newblock page 99090V, 07 2016.

\bibitem{Perrin2003}
Marshall~D. Perrin, Anand Sivaramakrishnan, Russell~B. Makidon, Ben~R.
  Oppenheimer, and James~R. Graham.
\newblock {The structure of high strehl ratio point-spread functions}.
\newblock {\em Astrophys. J.}, 596:702--712, 2003.

\bibitem{Poyneer16}
Lisa~A. Poyneer, David~W. Palmer, Bruce Macintosh, Dmitry Savransky, Naru
  Sadakuni, Sandrine Thomas, Jean-Pierre V\'{e}ran, Katherine~B. Follette,
  Alexandra~Z. Greenbaum, S.~Mark Ammons, Vanessa~P. Bailey, Brian Bauman,
  Andrew Cardwell, Daren Dillon, Donald Gavel, Markus Hartung, Pascale Hibon,
  Marshall~D. Perrin, Fredrik~T. Rantakyr\"{o}, Anand Sivaramakrishnan, and
  Jason~J. Wang.
\newblock Performance of the gemini planet imager's adaptive optics system.
\newblock {\em Appl. Opt.}, 55(2):323--340, Jan 2016.

\bibitem{Pueyo2012}
Laurent Pueyo, Justin~R. Crepp, Gautam Vasisht, Douglas Brenner, Ben~R.
  Oppenheimer, Neil Zimmerman, Sasha Hinkley, Ian Parry, Charles Beichman,
  Lynne Hillenbrand, Lewis~C. Roberts, Richard Dekany, Mike Shao, Rick Burruss,
  Antonin Bouchez, Jenny Roberts, and R{\'{e}}mi Soummer.
\newblock {APPLICATION} {OF} a {DAMPED} {LOCALLY} {OPTIMIZED} {COMBINATION}
  {OF} {IMAGES} {METHOD} {TO} {THE} {SPECTRAL} {CHARACTERIZATION} {OF} {FAINT}
  {COMPANIONS} {USING} {AN} {INTEGRAL} {FIELD} {SPECTROGRAPH}.
\newblock {\em The Astrophysical Journal Supplement Series}, 199(1):6, feb
  2012.

\bibitem{Rajan2017}
Abhijith Rajan, Julien Rameau, Robert J.~De Rosa, Mark~S. Marley, James~R.
  Graham, Bruce Macintosh, Christian Marois, Caroline Morley, Jennifer
  Patience, Laurent Pueyo, Didier Saumon, Kimberly Ward-Duong, S.~Mark Ammons,
  Pauline Arriaga, Vanessa~P. Bailey, Travis Barman, Joanna Bulger, Adam~S.
  Burrows, Jeffrey Chilcote, Tara Cotten, Ian Czekala, Rene Doyon, Gaspard
  Duch{\^{e}}ne, Thomas~M. Esposito, Michael~P. Fitzgerald, Katherine~B.
  Follette, Jonathan~J. Fortney, Stephen~J. Goodsell, Alexandra~Z. Greenbaum,
  Pascale Hibon, Li-Wei Hung, Patrick Ingraham, Mara Johnson-Groh, Paul Kalas,
  Quinn Konopacky, David Lafreni{\`{e}}re, James~E. Larkin, J{\'{e}}r{\^{o}}me
  Maire, Franck Marchis, Stanimir Metchev, Maxwell~A. Millar-Blanchaer,
  Katie~M. Morzinski, Eric~L. Nielsen, Rebecca Oppenheimer, David Palmer,
  Rahul~I. Patel, Marshall Perrin, Lisa Poyneer, Fredrik~T. Rantakyrö,
  Jean-Baptiste Ruffio, Dmitry Savransky, Adam~C. Schneider, Anand
  Sivaramakrishnan, Inseok Song, R{\'{e}}mi Soummer, Sandrine Thomas, Gautam
  Vasisht, J.~Kent Wallace, Jason~J. Wang, Sloane Wiktorowicz, and Schuyler
  Wolff.
\newblock Characterizing 51 eri b from 1 to 5 $\mu$m: A partly cloudy
  exoplanet.
\newblock {\em The Astronomical Journal}, 154(1):10, jun 2017.

\bibitem{roggemann1998}
Michael~C Roggemann and David~J Lee.
\newblock Two-deformable-mirror concept for correcting scintillation effects in
  laser beam projection through the turbulent atmosphere.
\newblock {\em Applied Optics}, 37(21):4577--4585, 1998.

\bibitem{Seo2018}
Byoung-Joon Seo, Fang Shi, Bala Balasubramanian, Eric Cady, Brian Gordon, Brian
  Kern, Raymond Lam, David Marx, Dwight Moody, Richard Muller, Keith Patterson,
  Ilya Poberezhskiy, Camilo~Mejia Prada, A.J.~Eldorado Riggs, John Trauger, and
  Daniel Wilson.
\newblock Hybrid lyot coronagraph for wfirst: high contrast testbed
  demonstration in flight-like low flux environment.
\newblock volume 10698, pages 10698 -- 10698 -- 12, 2018.

\bibitem{Sivaramakrishnan2002}
Anand Sivaramakrishnan, James~P. Lloyd, Philip~E. Hodge, and Bruce~A.
  Macintosh.
\newblock Speckle decorrelation and dynamic range in speckle
  noise{\textendash}limited imaging.
\newblock {\em The Astrophysical Journal}, 581(1):L59--L62, dec 2002.

\bibitem{Tallis2018}
Melisa Tallis, Vanessa~P. Bailey, Bruce Macintosh, Jeffrey~K. Chilcote, Lisa~A.
  Poyneer, Jean-Baptiste Ruffio, Thomas~L. Hayward, and Dmitry Savransky.
\newblock Air, telescope, and instrument temperature effects on the gemini
  planet imager’s image quality.
\newblock volume 10703, pages 10703 -- 10703 -- 6, 2018.

\bibitem{Tatarski1961}
V.I. Tatarski.
\newblock {\em Wave Propagation in a Turbulent Medium}.
\newblock Dover books on physics and mathematical physics. Dover, 1961.

\bibitem{Troy2006}
M.~{Troy}, I.~{Crossfield}, G.~{Chanan}, P.~{Dumont}, J.~J. {Green}, and
  B.~{Macintosh}.
\newblock {Effects of diffraction and static wavefront errors on high-contrast
  imaging from the thirty meter telescope}.
\newblock In {\em Society of Photo-Optical Instrumentation Engineers (SPIE)
  Conference Series}, volume 6272 of {\em procspie}, page 62722C, June 2006.

\bibitem{pyklip}
J.~J. {Wang}, J.-B. {Ruffio}, R.~J. {De Rosa}, J.~{Aguilar}, S.~G. {Wolff}, and
  L.~{Pueyo}.
\newblock {pyKLIP: PSF Subtraction for Exoplanets and Disks}.
\newblock Astrophysics Source Code Library, June 2015.

\bibitem{Wang2018}
Jason~J. Wang, James~R. Graham, Rebekah Dawson, Daniel Fabrycky, Robert J.~De
  Rosa, Laurent Pueyo, Quinn Konopacky, Bruce Macintosh, Christian Marois,
  Eugene Chiang, S.~Mark Ammons, Pauline Arriaga, Vanessa~P. Bailey, Travis
  Barman, Joanna Bulger, Jeffrey Chilcote, Tara Cotten, Rene Doyon, Gaspard
  Duch{\^{e}}ne, Thomas~M. Esposito, Michael~P. Fitzgerald, Katherine~B.
  Follette, Benjamin~L. Gerard, Stephen~J. Goodsell, Alexandra~Z. Greenbaum,
  Pascale Hibon, Li-Wei Hung, Patrick Ingraham, Paul Kalas, James~E. Larkin,
  J{\'{e}}r{\^{o}}me Maire, Franck Marchis, Mark~S. Marley, Stanimir Metchev,
  Maxwell~A. Millar-Blanchaer, Eric~L. Nielsen, Rebecca Oppenheimer, David
  Palmer, Jennifer Patience, Marshall Perrin, Lisa Poyneer, Abhijith Rajan,
  Julien Rameau, Fredrik~T. Rantakyrö, Jean-Baptiste Ruffio, Dmitry Savransky,
  Adam~C. Schneider, Anand Sivaramakrishnan, Inseok Song, Remi Soummer,
  Sandrine Thomas, J.~Kent Wallace, Kimberly Ward-Duong, Sloane Wiktorowicz,
  and Schuyler Wolff.
\newblock Dynamical constraints on the {HR} 8799 planets with {GPI}.
\newblock {\em The Astronomical Journal}, 156(5):192, oct 2018.

\bibitem{Wu18}
Chensheng Wu, Jonathan Ko, John~R. Rzasa, Daniel~A. Paulson, and Christopher~C.
  Davis.
\newblock Phase and amplitude beam shaping with two deformable mirrors
  implementing input plane and fourier plane phase modifications.
\newblock {\em Appl. Opt.}, 57(9):2337--2345, Mar 2018.

\bibitem{Zhou10}
Ping Zhou and James~H. Burge.
\newblock Analysis of wavefront propagation using the talbot effect.
\newblock {\em Appl. Opt.}, 49(28):5351--5359, Oct 2010.

\end{thebibliography}

\newpage
\renewcommand{\thefigure}{A\arabic{figure}}
\setcounter{figure}{0}

\section*{Appendix A: Propagation through the Atmosphere}
Tartarski\cite{Tatarski1961} has shown that the fluctuations in the optical index of refraction in three dimensions for a Kolmogorov turbulence spectrum follow the form
\begin{equation}
\Phi_N(\kappa, z) = 0.033 C_N^2(z)\kappa^{-11/3}
\end{equation}
where $C_N^2$ is the index of refraction structure constant and and $\kappa = 2\pi/l$ is the spatial wavevector for an eddy of size $l$. Here, we use the standard Kolmogorov power spectrum, which is fractally self-similar at all length scales, although it is in principle simple to extend this model to a Von-Karman spectrum by attenuating the power above and below the outer and inner scales. From the square root of the power spectrum, we can find the fluctuations from the inverse Fourier transform according to Johansson\cite{Johansson1994} with

\begin{equation}
\delta N (\vec{x},z) = \textrm{Re}\Big[\mathcal{F}^{-1}\Big(\xi(\vec{\kappa},z)\sqrt{\Phi_N(\kappa,z)}\Big)\Big]
\end{equation}
where $\delta N$ are the fluctuations of the index of refraction from unity in parts per million, $\xi$ is a zero-mean unit-variance complex hermitian Gaussian noise process, and $\mathcal{F}^{-1}$ is the unnormalized inverse discrete Fourier transform given by
\begin{equation}
\eta_{\textrm{ab}} = \mathcal{F}^{-1}(\tilde{\eta}_{\textrm{pq}}) = \sum_{p=0}^{P-1} \sum_{q=0}^{Q-1} \tilde{\eta}_{\textrm{pq}} \exp\Big[2\pi i\Big(\frac{pa}{P} + \frac{qb}{Q}\Big)\Big]
\end{equation}
for a discrete array of size $P\times Q$ with $P,Q \in \mathbb{N}$. The discrete indices $p,a \in {0,1, ..., P-1}$ and $q,b \in {0,1, ..., Q-1}$ exist in Fourier and configuration space, respectively. The corresponding forward Fourier transform simply includes negation in the exponent, and we have to pay careful attention the the normalization factor used by a routine such as np.fft.fft2, which includes a normalization of $\frac{1}{PQ}$ on the inverse transform, but no normalization on the forward transform by default.

The optical path length of a wavefront traversing a turbulent layer in the atmosphere from zenith can be found to first order by integrating the index of refraction over the thickness of the layer, and the accumulated phase is simply the wavevector of the ray $k = 2\pi/\lambda$ times the optical path length.
\begin{equation}
\phi_i(\vec{x}) = k\int_{z_i}^{z_i+\Delta z_i} n(\vec{x},z) \textrm{d}z = k n(\vec{x},z)\Delta z_i
\end{equation}
Here $\vec{x} = (x, y)$ is the coordinate system in the aperture at $z=0$, $\Delta z_i$ is the range of altitudes relevant to the turbulent layer at altitude $z_i$, and the baseline index of refraction of the atmosphere can be approximated\cite{Hardy1998} with 
\begin{equation}
N \equiv (n-1)10^6 \approx 77.6 \frac{\rho}{T}
\end{equation}
where $\rho$ and $T$ are the pressure (in millibars or equivalently hPa) and temperature (in Kelvin) of the atmosphere for a particular altitude. To obtain a model for the index of refraction as a function of altitude, one can model the pressure as a decaying exponential with a scale height given by the local surface gravity $g$, the mean molecular mass $M$ of the atmosphere, the ideal gas constant $R$, and an assumed isothermal uniform temperature of the surface $T$, with

\begin{equation}
\rho(z) = \rho_0 e^{\frac{-Mgz}{RT}}
\end{equation}
where $\rho_0$ is the atmospheric pressure at sea level. An atmospheric temperature profile as a function from altitude can be determined empirically, or the values given in the GFS can be used, but small fluctuations in T hardly affect the end value of the index of refraction, compared to the pressure, which dominates.

For non-zenith observations an additional term of $\sec{\zeta}$ where $\zeta$ is the zenith angle should be included in the integral in (40) to account for additional atmospheric depth. When the accumulated phase on the aperture is very large, we can subtract off the average phase, which is equivalent to removing the piston term from a Zernike Polynomial.\cite{MalesGuyon17}

In order to account for scintillation, each turbulent layer must be propagated according to the angular spectrum rule derived at the beginning of this paper. In order to make this simulation numerically tractable, the propagation through the turbulent phase screen is discretized into two steps, one where the phase is first accumulated according to the entire thickness of the layer at the start, and then second where the wave free space propagates the entire distance of the layer. For an infinite number of layers this assumption should recover the true propagation and indeed we are in a regime where the layer thickness is relatively small compared to the total propagation distance. A shorthand summary rule for the angular spectrum propagation is that the complex illumination at propagation distance $z$ is related to the complex illumination at the origin with

\begin{equation}
    u(z) = \mathcal{F}^{-1}[H(z) \mathcal{F}[u]]
\end{equation}
where $H$ is the free space propagation transfer function which is implicitly also a function of the particular modes $k$ being propagated. With the complex illumination given at the aperture by the previous description, we invoke the Taylor frozen-flow hypothesis, which requires that the timescale for turbulence is much greater than the time delay with which the AO system will respond. For our simulation, this means that the fluctuations in the field of view simply propagate by translations due to the wind velocity, which can be expressed by
\begin{equation}
\delta N(\vec{x}+\vec{v}(z)\tau,t_0+\tau) = \delta N(\vec{x}, t_0)
\end{equation}
where $\vec{v}(z)$ is the wind velocity at altitude $z$, which is assumed to lie only in the plane at altitude with no vertical component, $t_0$ is an particular instant in time, and $\tau$ is the total time delay for the adaptive optics system to respond to a measurement from the wavefront sensor. We also assume a perfect noiseless wavefront sensor and deformable mirror with only a time lag or servo lag error for an ideal open-loop AO simulation. The expression for the compensated phase in the aperture is a new complex illumination with the of the amplitude errors from the current timestep and the phase given by two subtracted phases, one from the current timestep and one from the previous, which is our AO correction
\begin{equation}
\phi_{\textrm{AO}} = \phi(\vec{x},t_0) - \phi(\vec{x},t_0-\tau)
\end{equation}
where we implicitly have included the the contributions from $L$ turbulent layers at various altitudes $z$ with a flat interpolation scheme for the structure constant. From the compensated phase on the aperture, we can obtain the final image's intensity distribution with an inverse Fourier transform by assuming the telescope focus operates in a Fraunhofer diffraction limit, so that electric field distribution in the image plane is the inverse Fourier transform of the aperture function\cite{Hecht2002}.

\begin{equation}
I(\alpha, \beta) = | \langle \mathcal{F}^{-1}(\mathcal{A}U_{\textrm{AO}})\rangle |^2
\end{equation}
Here $\alpha$, $\beta$ are the coordinates in the image plane, $U_{\textrm{AO}}$ is our AO corrected complex illumination which includes amplitude errors from scintillation, $\mathcal{A}$ is the aperture function with the Blackman window apodization \cite{Blackman}, parameterized radially from the center with $r^2 = x^2 + y^2$

\begin{equation}
    \mathcal{A}(r) = \frac{1 - \gamma}{2} - \frac{1}{2}\cos\Big(\frac{2\pi(r-D/2)}{D}\Big) + \frac{\gamma}{2}\cos\Big(\frac{2\pi(r-D/2}{D}\Big)
\end{equation}
with an aperture diameter of $D = 8$ m, a falloff of $\gamma = .16$, and the brackets denote time average over the whole length of the simulation. The Blackman apodization simulates a crude coronagraph and dampens the high order airy rings, whose final intensity in the image plane can swamp the effect of the scintillation halo.

\newpage
\section*{Appendix B: The Relationship between the Sky and the Ground}
The Back of the Telescope (BT) Plane is the simplest way to imagine the relationship between an image on the sky and its orientation relative to the ground. Suppose you have a DSLR on a tripod, or a multi-million dollar telescope with an Alt-Az tracking system. Either way\footnote{It is worth noting that the validity of this analogy, as well as is necessary to implement Angular Differential Imaging, a post-processing technique for combining multiple exposures while the target star moves through the zenith, that GPI operates in a fixed parallactic orientation, with the instrument derotator disabled, so that GPI is fixed with respect to the telescope orientation, which is uncommon.}, your imaging device is pointed at the celestial sphere along the line of sight vector 
\begin{equation}
\hat{r} = \langle\cos(el)\cos(az), -\cos(el)\sin(az), \sin(el)\rangle
\end{equation}
Where we have assumed the convention of the positive x-axis pointing North, and the positive y-axis pointing West. This conveniently sets up the positive z-axis to point towards Zenith, as it should. Azimuth is measured from North opening towards the East, and elevation is measured from the horizon upwards. See Figure A1 for an illustration.

\begin{figure}[h!]
\centering
\includegraphics[width=.5\textwidth]{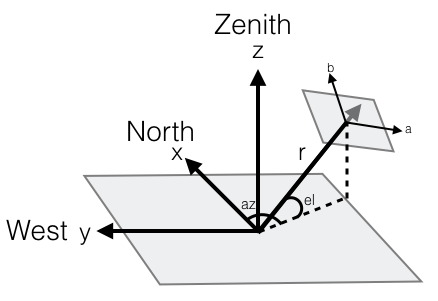}
\caption{Diagram of coordinates used to orient images on the sky demonstrating the relationship between the ground plane and the BT plane.}
\end{figure}

With such conventions laid out, it becomes easy to identify the location of the image plane on the sky, as it must be perpendicular to the line of sight. Since there are infinitely many such planes, we will use the convention
\begin{eqnarray}
\hat{a} &=& \langle - \sin(az), - \cos(az), 0 \rangle \\
\hat{b} &=& \langle-\sin(el)\cos(az), \sin(el)\sin(az), \cos(el)\rangle
\end{eqnarray}
So that one can think of $\hat{a}$ as pointing in the direction of increasing Azimuth, and $\hat{b}$ pointing towards increasing Elevation. It is left to the reader to show that $\hat{a} \cdot \hat{b} = 0$, and that $\hat{a} \times \hat{r} = \hat{b}$ to verify the orthogonality of these unit vectors as a coordinate system.

With this elaborate set up, it becomes easy to convert vectors in the image plane into vectors in full three dimensional space, and then project them onto the ground plane. Suppose we have a wind vector which appears in the image plane rotated $\psi$ from $\hat{a}$ counterclockwise. Such a wind vector is
\begin{equation}
\hat{w} = \cos(\psi)\hat{a} + \sin(\psi)\hat{b}
\end{equation}
However, we would instead like to know $\hat{w}(\hat{x},\hat{y},{\hat{z}})$. By algebraically substituting in our coordinate vectors $\hat{a}$, and $\hat{b}$ formulas in x,y,z space, we can arrive at an expression for the wind vector in x,y,z space in terms of $\psi$, $az$, and $el$. This is
\begin{align}
\hat{w} = \langle -\cos(\psi)\sin(az) - \sin(\psi)\sin(el)\cos(az), \nonumber\\
-\cos(\psi)\cos(az) + \sin(\psi)\sin(el)\sin(az), \nonumber\\ 
\sin(\psi)\cos(el)\rangle
\end{align}
With this done, we can easily project the vector onto the ground plane by simply removing the z-component. If we need to find the direction of this wind vector as an azimuth, we can use the following trick
\begin{equation}
\textrm{azimuth} =\Big(360\degree - \arctan 2\Big(\frac{w_y}{w_x}\Big)\Big) \% 360 \degree
\end{equation}
Where $w_x$, $w_y$ are the x and y components of the wind vector, respectively, \% is the modulo operator, and it is often convenient to use a smart operator like arctan2 to get the quadrant correct.

However, images in the GPIES are not simply oriented as in the BT plane, but rather can be arbitrarily arranged due to the complexities of post-processing. Fortunately for us, the orientation of each of the image has been previously calculated in celestial coordinates. These are represented as a CD Matrix, which describe how x and y in pixels for the image correspond to right ascension and declination. Using the local sidereal time of the image during the exposure, it is possible to convert coordinates in right ascension and declination to coordinates in azimuth and elevation, using

\begin{eqnarray}
\textrm{azimuth} &=& \Big(\arctan2\Big(\frac{\cos(\delta)\sin(h)}{\sin(\phi_0)\cos(\delta)\cos(h) - \cos(\phi_0)\sin(\delta)}\Big) + 180\Big) \% 360 \\
\textrm{elevation} &=& \arcsin(\sin(\phi_0)\sin(\delta) + \cos(\phi_0)\cos(\delta)\cos(h) \\
\end{eqnarray}
where $h = \theta_L - \alpha$ is the hour angle, $\theta_L$ is the local sidereal time in radians, $\phi_0$ is the local latitude, $\alpha$ is right ascension, $\delta$ is declination, and here we use the convention that azimuth starts at North and opens to the East. The modulo is there to handle overflow and the azimuth and elevation are the coordinates on the sky. Once these are calculated, we can orient images relative to the BT plane because $\hat{a}$ points towards increasing azimuth and $\hat{b}$ points towards increasing elevation.

\newpage
\section*{Appendix C: Calculating the 2nd order PSF}
For some electric field $U$ incident upon an aperture, described by
\begin{equation}
    U = A e^{iφ}
\end{equation}
and defining the change of case to represent the Fourier Transform
\begin{eqnarray}
    a &≡& \F [A] \\
    Φ &≡& \F [φ] \\ 
    u &≡& \F [U]
\end{eqnarray}
The PSF formed by such an electric field can be calculated with (Sivaramakrishnan et al. and Perrin et al.)
\begin{eqnarray}
    \PSF &=& |u|^2 = uu^* \\
    \PSF &≈& \PSF_0 + \PSF_1 + \PSF_2
\end{eqnarray}
Where the first few terms of the Taylor series are
\begin{eqnarray}
    \PSF_0 &=& aa^* \\
    \PSF_1 &=& 2 \textrm{Im}[a(a^* \star Φ^*)] \\
    \PSF_2 &=& (a\star Φ)(a^*\star Φ^*) - \frac{1}{2}[a(a^*\starΦ^*\starΦ^*) + a^*(a\starΦ\starΦ)]
\end{eqnarray}
In our one dimensional analytic model, the amplitude and phase that result after a time-delayed AO phase correction of an incoming scintillated wavefront are
\begin{eqnarray}
    A &=& 1 + \A\sin(2π(x-vt)/p) \\
    φ &=& P\sin(2π(x-Δ)/p)
\end{eqnarray}
where we have used the shorthand notations
\begin{eqnarray}
    \A &=& α\sin(2πz/z_T) \\
    P &=& α\cos(2πz/z_T)\sqrt{2}\sqrt{1 - \cos(2πvt/p)} \\
    Δ &=& -\frac{p}{2π} \arctan \Big[\frac{-\sin(2πvt/p)}{\cos(2πvt/p) - 1}\Big]
\end{eqnarray}
to simplify the expressions. In the following sections, we calculate $a$, $Φ$, and then use them to obtain PSF$_0$, PSF$_1$, PSF$_2$.

\subsection*{Calculating a}
Starting with the definition of $a$ and the definition of the Fourier Transform
\begin{equation}
    a = \F [A] = \int_{-∞}^{∞} [1 + \A \sin(2π/p(x-vt))]e^{-i2πfx}\d x
\end{equation}
and using the property that $\sin(x) = \frac{1}{2i}(e^{ix} - e^{-ix})$
\begin{eqnarray}
    a &=& \int_{-∞}^{∞} \Big[1 + \frac{\A}{2i}\Big(e^{i\frac{2π}{p}(x-vt)} - e^{-i\frac{2π}{p}(x-vt)}\Big)\Big]e^{-i2πfx}\d x \\
    a &=& \int_{-∞}^{∞} \Big[e^{-2πfx} + \frac{\A}{2i}\Big( e^{i2πx(1/p - f)}e^{-i\frac{2π}{p}vt} -  e^{-i2πx(1/p + f)}e^{i\frac{2π}{p}vt} \Big)\Big]\d x
\end{eqnarray}
along with the identity that $δ(x-α) = \int_{-∞}^{∞} e^{i2πf(x-α)}\d f$
\begin{equation}
    a = δ(f) + \frac{\A}{2i}δ(f-1/p)e^{-i\frac{2π}{p}vt} - \frac{\A}{2i}δ(f+1/p)e^{i\frac{2π}{p}vt}
\end{equation}

\subsection*{Calculating Φ}
Likewise
\begin{eqnarray}
    Φ = \F [φ] &=& \int_{-∞}^{∞} P\sin(2π/p(x-Δ))e^{-i2πfx}\d x \\
    &=& \int_{-∞}^{∞} \frac{P}{2i}\Big(e^{i\frac{2π}{p}(x-Δ)} - e^{-i\frac{2π}{p}(x-Δ)}\Big)e^{-i2πfx}\d x \\
    &=& \int_{-∞}^{∞} \frac{P}{2i}e^{i2πx(1/p - f)}e^{-i\frac{2π}{p}Δ} -\frac{P}{2i}e^{-i2πx(1/p - f)}e^{i\frac{2π}{p}Δ}\d x \\
    &=& \frac{P}{2i}δ(f-1/p)e^{-i\frac{2π}{p}Δ} -\frac{P}{2i}δ(f+1/p)e^{i\frac{2π}{p}Δ}
\end{eqnarray}

\subsection*{Calculating PSF$_0$}
Since $\PSF_0 = aa^*$,
\begin{align}
    \PSF_0 = \Big[δ(f) + \frac{\A}{2i}δ(f-1/p)e^{-i\frac{2π}{p}vt} - \frac{\A}{2i}δ(f+1/p)e^{i\frac{2π}{p}vt} \Big] \times \nonumber \\
    \Big[ δ(f) + \frac{\A}{-2i}δ(f-1/p)e^{i\frac{2π}{p}vt} - \frac{\A}{-2i}δ(f+1/p)e^{-i\frac{2π}{p}vt} \Big]
\end{align}
And using the mathematically non-rigorous (though conceptually clear) notions that $[δ(f-α)]^2 = δ(f-α)$ and $δ(f-α)δ(f-β) = 0$ when $α≠β$, then the cross terms cancel and the complex terms go to their moduli squared and the equation simplifies to
\begin{equation}
    \PSF_0 = δ(f) + \frac{\A^2}{4}\Big[δ(f-1/p) + δ(f+1/p)\Big].
\end{equation}

\subsection*{Calculating PSF$_1$}
Starting with the definition of PSF$_1$
\begin{equation}
    \PSF_1 = 2\textrm{Im}[a(a^*\starΦ^*)],
\end{equation}
and the definition of the convolution operation
\begin{equation}
    (a^* \star Φ^*)(f) = \int_{-∞}^{∞} a^*(f')Φ^*(f-f')\d f' ,
\end{equation}
\begin{align}
    (a^* \star Φ^*)(f) = \int_{-∞}^{∞} \Big[ δ(f') - \frac{\A}{2i}δ(f'-1/p)e^{i\frac{2π}{p}vt} + \frac{\A}{2i}δ(f'+1/p)e^{-i\frac{2π}{p}vt}\Big] \times  \nonumber\\
    \Big[\frac{P}{-2i}δ((f-f')-1/p)e^{i\frac{2π}{p}Δ} +\frac{P}{2i}δ((f-f')+1/p)e^{-i\frac{2π}{p}Δ}\Big]
\end{align}
and using the sifting property $\int_{-∞}^{∞} f(t) δ(t-τ)\d t = f(τ)$ six times, once for each term,
\begin{align}
    \int_{-∞}^{∞} δ(f') \frac{P}{-2i}δ((f-f')-1/p)e^{i\frac{2π}{p}Δ} \d f' &= \frac{P}{-2i}e^{i\frac{2π}{p}Δ}δ(f-1/p) \\
    \int_{-∞}^{∞} δ(f') \frac{P}{2i}δ((f-f')+1/p)e^{-i\frac{2π}{p}Δ} \d f' &= \frac{P}{2i}e^{-i\frac{2π}{p}Δ}δ(f+1/p) \\
    \int_{-∞}^{∞} \frac{\A}{-2i}δ(f'-1/p)e^{i\frac{2π}{p}vt} \frac{P}{-2i}δ((f-f')-1/p)e^{i\frac{2π}{p}Δ} \d f' &= \frac{-\A P}{4}e^{i\frac{2π}{p}(vt +Δ)}δ(f-2/p) \\
    \int_{-∞}^{∞} \frac{\A}{-2i}δ(f'-1/p)e^{i\frac{2π}{p}vt} \frac{P}{2i}δ((f-f')+1/p)e^{-i\frac{2π}{p}Δ} \d f' &= \frac{\A P}{4}e^{i\frac{2π}{p}(vt -Δ)} δ(-f) \\
    \int_{-∞}^{∞} \frac{\A}{2i}δ(f'+1/p)e^{-i\frac{2π}{p}vt} \frac{P}{-2i}δ((f-f')-1/p)e^{i\frac{2π}{p}Δ} \d f' &= \frac{\A P}{4}e^{-i\frac{2π}{p}(vt -Δ)} δ(f) \\
    \int_{-∞}^{∞} \frac{\A}{2i}δ(f'+1/p)e^{-i\frac{2π}{p}vt} \frac{P}{2i}δ((f-f')+1/p)e^{-i\frac{2π}{p}Δ} \d f' &= \frac{-\A P}{4}e^{-i\frac{2π}{p}(vt +Δ)} δ(f + 2/p) .
\end{align}
Here it is worth noting that the $δ$ function is even and so obeys the relationship that $δ(-x) = δ(x)$. Furthermore, factoring out common terms
\begin{align}\label{astarphistar}
    (a^* \star Φ^*) = \frac{\A P}{4}\Big[e^{i\frac{2π}{p}(vt -Δ)} δ(f) &-e^{i\frac{2π}{p}(vt +Δ)}δ(f-2/p) \nonumber \\
    e^{-i\frac{2π}{p}(vt -Δ)} δ(f) &-e^{-i\frac{2π}{p}(vt +Δ)}δ(f + 2/p) \Big] \nonumber \\
    + \frac{P}{2i}\Big[e^{-i\frac{2π}{p}Δ}δ(f+1/p) &- e^{i\frac{2π}{p}Δ}δ(f-1/p)\Big] .
\end{align}
Multiplying by $a$, only the cross terms with $δ(f)$, $δ(f-1/p)$, and $δ(f+1/p)$ remain
\begin{align}
    (a(a^* \star Φ^*))(f) = \frac{\A P}{4}\Big[e^{i\frac{2π}{p}(vt -Δ)} δ(f) 
     + e^{-i\frac{2π}{p}(vt -Δ)} δ(f) \Big] \nonumber \\
     + \frac{\A P}{4 i^2}\Big[ -δ(f-1/p)e^{-i\frac{2π}{p}vt}e^{i\frac{2π}{p}Δ} - δ(f+1/p)e^{i\frac{2π}{p}vt}e^{-i\frac{2π}{p}Δ} \Big]
\end{align}
factoring again,
\begin{equation}
    (a(a^* \star Φ^*))(f) = \frac{\A P}{4}\Big[e^{i\frac{2π}{p}(vt -Δ)}\Big(δ(f) + d(f+1/p)\Big) + e^{-i\frac{2π}{p}(vt -Δ)}\Big(δ(f) + δ(f-1/p)\Big) \Big]
\end{equation}
And taking twice the imaginary part, using Im$[e^{iαx}] = \sin(αx)$
\begin{align}
    2\textrm{Im}[(a(a^* \star Φ^*))](f) = \frac{\A P}{2}\Big[\sin\Big(\frac{2π}{p}(vt -Δ)\Big)\Big(δ(f) + δ(f+1/p)\Big) \nonumber \\
    + \sin\Big(-\frac{2π}{p}(vt -Δ)\Big)\Big(δ(f) + δ(f-1/p)\Big) \Big]
\end{align}
However, since sine is odd and $\sin(-x) = -\sin(x)$, the term with $δ(f)$ cancels and
\begin{equation}
    \PSF_1 = 2\textrm{Im}[(a(a^* \star Φ^*))](f) = \frac{\A P}{2}\sin\Big(\frac{2π}{p}(vt -Δ)\Big) \Big[δ(f+1/p) - δ(f-1/p)\Big]
\end{equation}
and as we can see from the minus sign it is antisymmetric.

\subsection*{Calculating PSF$_2$}
Since
\begin{equation}
     \PSF_2 = (a\star Φ)(a^*\star Φ^*) - \frac{1}{2}[a(a^*\starΦ^*\starΦ^*) + a^*(a\starΦ\starΦ)],
\end{equation}
we start first by calculating the first term $\PSF_{2,\textrm{halo}} = (a\star Φ)(a^*\star Φ^*) $. Since we have already calculated $(a^*\star Φ^*)$ in the previous section (see equation \ref{astarphistar}), we only need to find $(a \star Φ)$  . While, in general $(g^*\star h^*) ≠ (g\star h)^*$ for generic complex functions $g$ and $h$, here the complex conjugate does not modify the $δ$ function, which has no imaginary part, and therefore does not change the convolution integral, as it is the only part of the function dependent on the variable of integration $f'$. Since the complex conjugate only modifies the multiplicative constants with $i$ and the phasors $e^{iα}$, we can simply take the complex conjugate of (\ref{astarphistar}) and arrive at the answer.
\begin{align}
    (a\star Φ) = \frac{\A P}{4}\Big[e^{-i\frac{2π}{p}(vt -Δ)} δ(f) &-e^{-i\frac{2π}{p}(vt +Δ)}δ(f-2/p) \nonumber \\
    e^{+i\frac{2π}{p}(vt -Δ)} δ(f) &-e^{+i\frac{2π}{p}(vt +Δ)}δ(f + 2/p) \Big] \nonumber \\
    + \frac{P}{-2i}\Big[e^{+i\frac{2π}{p}Δ}δ(f+1/p) &- e^{-i\frac{2π}{p}Δ}δ(f-1/p)\Big] .
\end{align}
This convenient fact means that when we multiply the two terms together, nearly all of the complex phasors will go to their moduli squared, aka the constant $1$, simplifying the expression very nicely. However, since cross terms from both of the $δ(f)$ terms will exist, one with each $+$ and $-$ in the exponent, two of them cancel but two of them do not, and their exponents add up to having $4π$ terms.
\begin{align}
    (a\star Φ)(a^*\star Φ^*)  = \frac{\A^2 P^2}{16}\Big[\Big(2 + e^{+i\frac{4π}{p}(vt -Δ)} + e^{-i\frac{4π}{p}(vt -Δ)} \Big)δ(f) \nonumber \\
    + δ(f + 2/p) + δ(f - 2/p)\Big] + \frac{P^2}{4}\Big[δ(f+1/p) + δ(f-1/p)\Big]
\end{align}
These two exponents can be combined with the identity that $e^{iαx} + e^{-iαx} = 2 \cos(αx)$
\begin{align}
    \PSF_{2,\textrm{halo}} = \frac{\A^2 P^2}{16}\Big[2\Big(1 + \cos\Big(\frac{4π}{p}(vt -Δ)\Big)\Big)δ(f)
    + δ(f + 2/p) + δ(f - 2/p)\Big] \nonumber \\ 
    + \frac{P^2}{4}\Big[δ(f+1/p) + δ(f-1/p)\Big]
\end{align}

To find PSF$_{2,\textrm{strehl}}$, the second second-order term, we must perform an additional convolution, this time, with twelve terms.
\begin{equation}
     a^*\starΦ^*\starΦ^* = \int_{-∞}^{∞}  (a^*\starΦ^*)(f') Φ^*(f-f') \d f' 
\end{equation}
\begin{align}
     & = \int_{-∞}^{∞} \Big\{  \frac{\A P}{4}\Big[e^{i\frac{2π}{p}(vt -Δ)} δ(f') + e^{-i\frac{2π}{p}(vt -Δ)}δ(f') - e^{i\frac{2π}{p}(vt +Δ)}δ(f'-2/p)  - e^{-i\frac{2π}{p}(vt +Δ)}δ(f' + 2/p) \Big] \nonumber \\
    & + \frac{P}{2i}\Big[e^{-i\frac{2π}{p}Δ}δ(f'+1/p) - e^{i\frac{2π}{p}Δ}δ(f'-1/p)\Big] \Big\} \times \frac{P}{-2i}\Big[e^{i\frac{2π}{p}Δ}δ((f-f')-1/p) - e^{-i\frac{2π}{p}Δ}δ((f-f')+1/p)\Big] \d f'
\end{align}
\begin{align}
    \int_{-∞}^{∞} \frac{\A P}{4}e^{i\frac{2π}{p}(vt -Δ)}δ(f')\times \frac{P}{-2i}e^{i\frac{2π}{p}Δ}δ((f-f')-1/p) \d f' &= \frac{-\A P^2}{8i}e^{i\frac{2π}{p}vt}δ(f-1/p) \\
    \int_{-∞}^{∞} \frac{\A P}{4}e^{i\frac{2π}{p}(vt -Δ)}δ(f')\times
    \frac{P}{2i} e^{-i\frac{2π}{p}Δ}δ((f-f')+1/p) \d f' &= \frac{\A P^2}{8i}e^{i\frac{2π}{p}(vt -2Δ)}δ(f + 1/p) \\
    \int_{-∞}^{∞} \frac{\A P}{4}e^{-i\frac{2π}{p}(vt -Δ)}δ(f')\times \frac{P}{-2i}e^{i\frac{2π}{p}Δ}δ((f-f')-1/p) \d f' &= \frac{-\A P^2}{8i}e^{-i\frac{2π}{p}(vt -2Δ)}δ(f-1/p) \\
    \int_{-∞}^{∞} \frac{\A P}{4}e^{-i\frac{2π}{p}(vt -Δ)}δ(f')\times
    \frac{P}{2i} e^{-i\frac{2π}{p}Δ}δ((f-f')+1/p) \d f' &= \frac{\A P^2}{8i}e^{-i\frac{2π}{p}vt}δ(f+1/p)\\
    \int_{-∞}^{∞} \frac{-\A P}{4}e^{i\frac{2π}{p}(vt +Δ)}δ(f'-2/p)\times \frac{P}{-2i}e^{i\frac{2π}{p}Δ}δ((f-f')-1/p) \d f' &= \frac{\A P^2}{8i}e^{i\frac{2π}{p}(vt +2Δ)}δ(f'-3/p) \\
    \int_{-∞}^{∞} \frac{-\A P}{4}e^{i\frac{2π}{p}(vt +Δ)}δ(f'-2/p)\times
    \frac{P}{2i} e^{-i\frac{2π}{p}Δ}δ((f-f')+1/p) \d f' &= \frac{-\A P^2}{8i}e^{i\frac{2π}{p}vt}δ(f'- 1/p)\\
    \int_{-∞}^{∞} \frac{-\A P}{4}e^{-i\frac{2π}{p}(vt +Δ)}δ(f' + 2/p)\times \frac{P}{-2i}e^{i\frac{2π}{p}Δ}δ((f-f')-1/p) \d f' &= \frac{\A P^2}{8i}e^{-i\frac{2π}{p}vt}δ(f + 1/p) \\
    \int_{-∞}^{∞} \frac{-\A P}{4}e^{-i\frac{2π}{p}(vt +Δ)}δ(f' + 2/p)\times
    \frac{P}{2i} e^{-i\frac{2π}{p}Δ}δ((f-f')+1/p) \d f' &= \frac{-\A P^2}{8i}e^{-i\frac{2π}{p}(vt +2Δ)}δ(f + 3/p) \\
    \int_{-∞}^{∞} \frac{P}{2i}e^{-i\frac{2π}{p}Δ}δ(f'+1/p)\times \frac{P}{-2i}e^{i\frac{2π}{p}Δ}δ((f-f')-1/p) \d f' &= \frac{P^2}{-4i^2}δ(f) \\
    \int_{-∞}^{∞} \frac{P}{2i}e^{-i\frac{2π}{p}Δ}δ(f'+1/p)\times
    \frac{P}{2i} e^{-i\frac{2π}{p}Δ}δ((f-f')+1/p) \d f' &= \frac{P^2}{4i^2}e^{-i\frac{4π}{p}Δ}δ(f+2/p) \\
    \int_{-∞}^{∞} \frac{P}{-2i}e^{i\frac{2π}{p}Δ}δ(f'-1/p)\times \frac{P}{-2i}e^{i\frac{2π}{p}Δ}δ((f-f')-1/p) \d f' &= \frac{P^2}{4i^2}e^{i\frac{4π}{p}Δ}δ(f-2/p) \\
    \int_{-∞}^{∞} \frac{P}{-2i}e^{i\frac{2π}{p}Δ}δ(f'-1/p)\times
    \frac{P}{2i} e^{-i\frac{2π}{p}Δ}δ((f-f')+1/p) \d f' &= \frac{P^2}{-4i^2}δ(f) \\
\end{align}
Multiplying by a, only the terms with $δ(f)$, $δ(f-1/p)$, and $δ(f+1/p)$ remain
\begin{align}
    a(a^*\starΦ^*\starΦ^*)  = \Big[ δ(f) + \frac{\A}{2i}δ(f-1/p)e^{-i\frac{2π}{p}vt} - \frac{\A}{2i}δ(f+1/p)e^{i\frac{2π}{p}vt}\Big] \times \Big\{\frac{\A P^2}{8i}\Big[ ... \Big] + \frac{P^2}{4}\Big[ ... \Big]\Big\} \\
    = \frac{P^2}{2}δ(f) + \frac{\A^2P^2}{16 i^2}\Big[\Big( -e^{i\frac{2π}{p}vt}-e^{-i\frac{2π}{p}(vt -2Δ)}-e^{i\frac{2π}{p}vt} \Big)e^{-i\frac{2π}{p}vt}δ(f-1/p)\nonumber \\
     - \Big(e^{i\frac{2π}{p}(vt -2Δ)} + e^{-i\frac{2π}{p}vt} + e^{-i\frac{2π}{p}vt} \Big)e^{i\frac{2π}{p}vt}δ(f+1/p)\Big] \\
     = \frac{P^2}{2}δ(f) + \frac{\A^2P^2}{16}\Big[\Big( 2 + e^{-i\frac{4π}{p}(vt -Δ)}\Big)δ(f-1/p) + \Big(2 + e^{i\frac{4π}{p}(vt -Δ)}\Big)δ(f+1/p)\Big]
\end{align}
where the $...$ represent the terms in (45) -- (57) after they have been properly factored.
It is a good exercise to perform the identical calculations with every single term complex conjugated, to find the corresponding second half of the second second-order term $a^*(a\starΦ\starΦ)$, which you may choose to do so on your own if you wish. Here we'll simply use the shortcut that since the $δ$ function is strictly real, $a^*(a\starΦ\starΦ) = [a(a^*\starΦ^*\starΦ^*)]^*$, to much more rapidly arrive at the answer.
\begin{align}
    a^*(a\starΦ\starΦ) = \frac{P^2}{2}δ(f) + \frac{\A^2P^2}{16}\Big[\Big( 2 + e^{i\frac{4π}{p}(vt -Δ)}\Big)δ(f-1/p) + \Big(2 + e^{-i\frac{4π}{p}(vt -Δ)}\Big)δ(f+1/p)\Big]
\end{align}
Adding these two together, and using $e^{iαx} + e^{-iαx} = 2 \cos(αx)$,
\begin{align}
    \PSF_{2,\textrm{strehl}} &= -\frac{1}{2}[a(a^*\starΦ^*\starΦ^*) + a^*(a\starΦ\starΦ)] \\
    &= -\frac{1}{2}\Big\{ P^2δ(f) + \frac{\A^2P^2}{16}\Big( 4 + 2\cos\Big(\frac{4π}{p}(vt -Δ)\Big) \Big)\Big[δ(f-1/p) + δ(f+1/p)\Big]\Big\}
\end{align}
In summary, we have calculated the following terms in the second order expansion of the AO-corrected and scintillated PSF.
\begin{flalign}
& \textrm{PSF}_0 = \delta(f) + \frac{\mathcal{A}^2}{4}\Big[\delta(f-1/p)+\delta(f+1/p)\Big] &\\
& \textrm{PSF}_1 =  \frac{-\mathcal{A}P}{2}\sin\Big(\frac{2\pi}{p}(vt-\Delta)\Big)\Big[\delta(f-1/p)-\delta(f+1/p) \Big] &\\
& \textrm{PSF}_{2,\textrm{halo}} = \frac{P^2}{4}\Big[δ(f+1/p) + δ(f-1/p)\Big]  + \frac{\A^2 P^2}{16}\Big[2\Big(1 + \cos\Big(\frac{4π}{p}(vt -Δ)\Big)\Big)δ(f)
    + δ(f + 2/p) + δ(f - 2/p)\Big]  &\\
& \textrm{PSF}_{2,\textrm{strehl}} = \frac{-P^2}{2}\delta(f) -\frac{\mathcal{A}^2P^2}{32}\Big(4 + 2\cos\Big(\frac{4\pi}{p}(vt-\Delta)\Big)\Big)\Big[\delta(f-1/p)+\delta(f+1/p)\Big].&
\end{flalign}
In general, obtaining an expression for the n$^\textrm{th}$-order term in the expansion will require performing an n-fold convolution, so calculations beyond the second order become increasingly tedious to perform.

\newpage
\section*{Additional Figures}
\begin{figure}[h!]
    \centering
    \includegraphics[width=\textwidth]{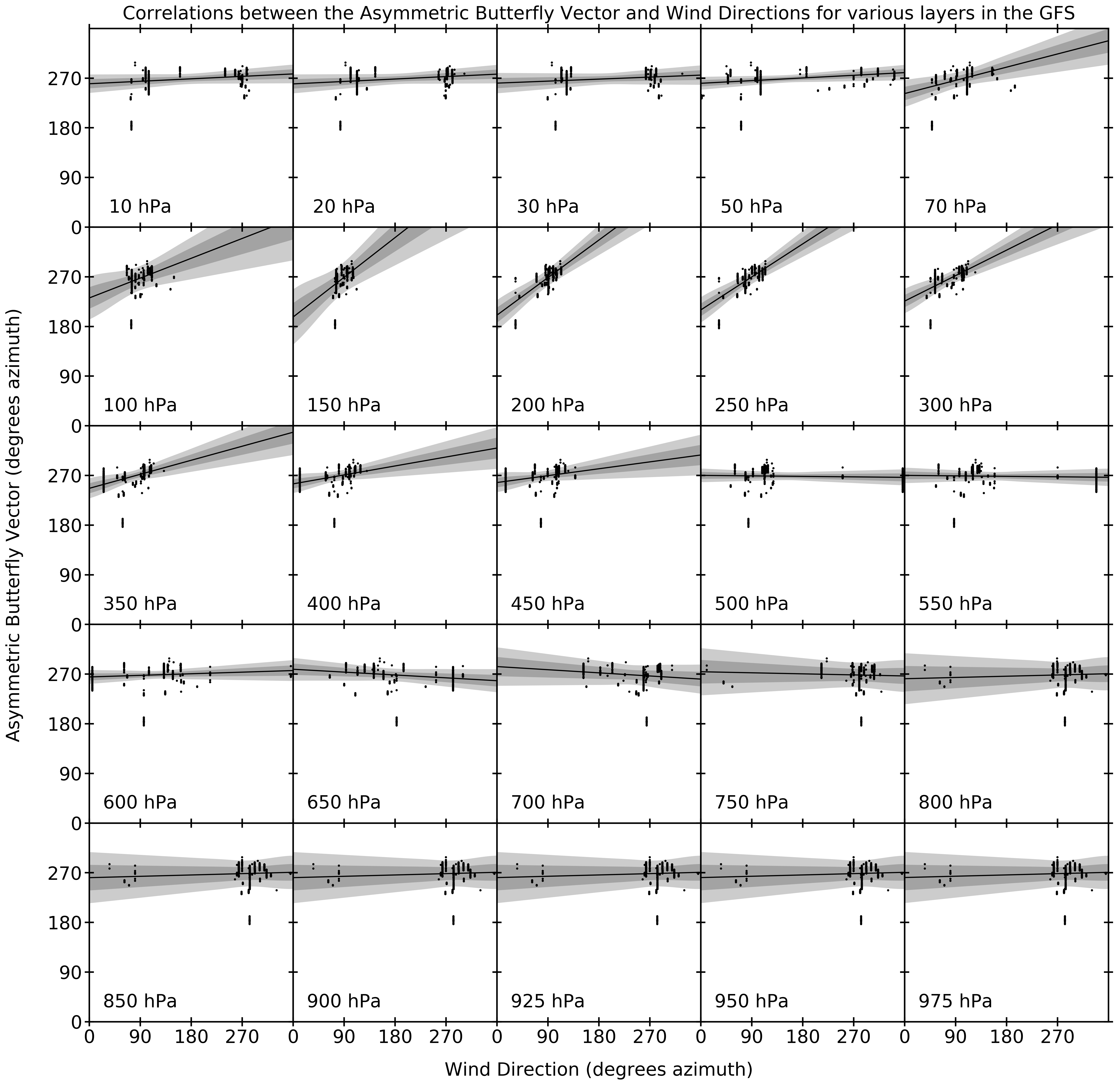}
    \caption{Correlations between the directions of the strong asymmetry of the image PSF and the wind direction for various wind layers in the NOAA GFS. Most wind layers do not exhibit significant correlation, with the exception of the layers around 100-250 hPa, which are the pressures corresponding the jet stream, at around 10-15 km of altitude.}
\end{figure}
\begin{figure}[h!]
    \centering
    \includegraphics[width=\textwidth]{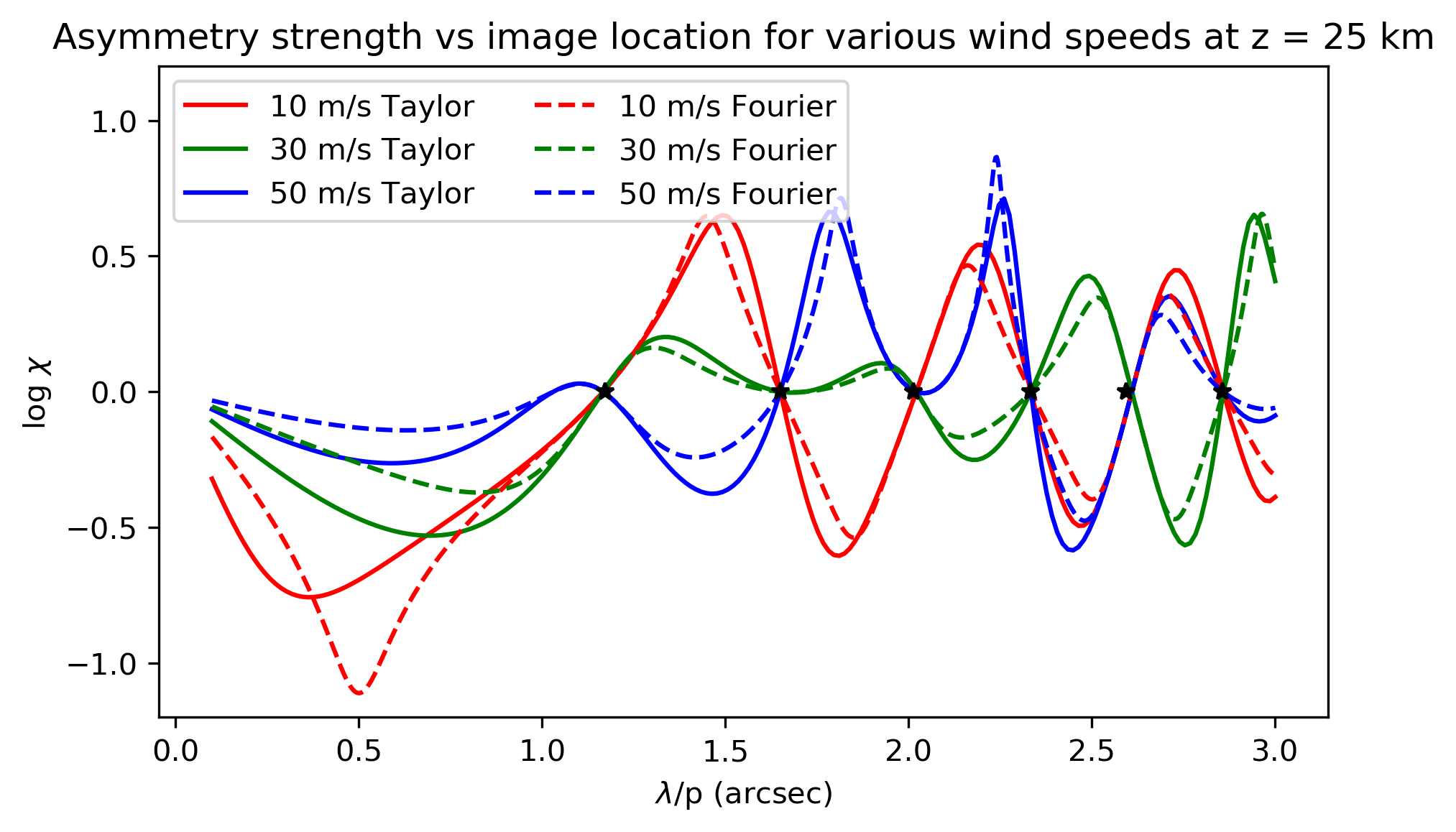}
    \caption{The log of the speckle asymmetry ratio for a single mode with an propagation distance of z = 25 km. Such layers in the atmosphere viewed at zenith are very sparse due to the exponential decline in pressure of the Earth's atmosphere, and so do not contribute significantly into observations done at zenith. For a jet stream layer at 15 km, a propagation distance of 25 km corresponds to telescope pointing elevation of 40 deg. When comparing this Figure to Figure 3, it becomes clear that modifying the propagation distance $z$ effectively changes the particular mode lengths the asymmetry occurs at to larger mode lengths, pushing the asymmetry to smaller separations, and nearer to the coronagraphic mask.}
\end{figure}
\begin{figure}[h!]
    \centering
    \includegraphics[width=\textwidth]{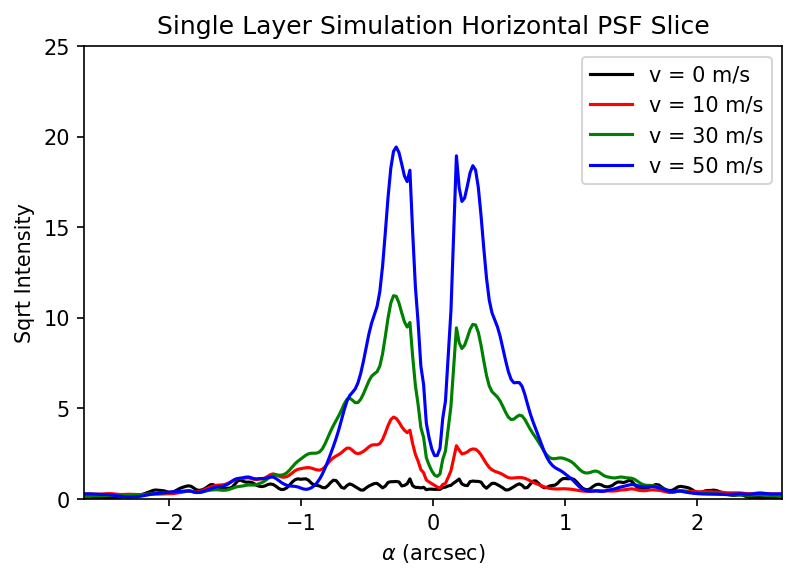}
    \caption{Horizontal slice through the center of the PSFs generated in the single layer atmospheric simulation, overlaid on top of each other for direct comparison. Here it is more explicitly visible as the wind speed decreases how both the total intensity of the halo decreases, as expected for better corrections, but in addition, the asymmetry becomes stronger at lower wind velocities, as expected from our model. For an infinitely fast AO system, the effective wind velocity is v = 0 m/s, which demonstrates the residual uncorrected amplitude error.}
\end{figure}

\end{document}